\documentclass[journal]{IEEEtran}
\ifCLASSINFOpdf
  \usepackage[pdftex]{graphicx}
  \graphicspath{../eps/}
  \DeclareGraphicsExtensions{.pdf,.jpeg,.png}
\else
  \usepackage[dvips]{graphicx}
  \graphicspath{{../eps/}}
  \DeclareGraphicsExtensions{.eps}
\fi
\usepackage{siunitx}
\usepackage[colorlinks=false]{hyperref}
\usepackage{color}
\usepackage{cite}
\usepackage{amsmath}
\usepackage{amssymb}
\usepackage{mathptmx}
\usepackage{epsfig}
\usepackage{times}
\usepackage{threeparttable}
\usepackage{stfloats}
\usepackage{amsmath,amssymb,amsfonts}
\usepackage{algorithmic}
\usepackage{textcomp}
\usepackage{array}

\usepackage{array}

  \usepackage[caption=false,font=footnotesize]{subfig}
\begin{document}
%
\title{Magnetoelectric Bio-Implants Powered and Programmed by a Single Transmitter for Coordinated Multisite Stimulation}
%
%
%

\author{
        Zhanghao Yu,~\IEEEmembership{Student Member,~IEEE,}
        Joshua C. Chen, 
        Yan He,~\IEEEmembership{Student Member,~IEEE,}
        Fatima~T.~Alrashdan,~\IEEEmembership{Student~Member,~IEEE,}
        Benjamin~W.~Avants,
        Amanda~Singer, 
        Jacob~T.~Robinson,~\IEEEmembership{Senior~Member,~IEEE,}
        and~Kaiyuan~Yang,~\IEEEmembership{Member,~IEEE}
\thanks{Manuscript received on}
\thanks{This work is based on research partially sponsored by the National Institutes of Health (NIH) under award U18EB029353, the National Science Foundation (NSF) under award ECCS-2023849, and 711 Human Performance Wing (HPW) and Defense Advanced Research Projects Agency (DARPA) under agreement number FA8650-21-1-7119. The U.S. Government is authorized to reproduce and distribute reprints for Governmental purposes notwithstanding any copyright notation thereon. (Corresponding Author: Kaiyuan Yang, kyang@rice.edu)}
\thanks{The views and conclusions contained herein are those of the authors and should not be interpreted as necessarily representing the official policies or endorsements either expressed or implied of 711 Human Performance Wing (HPW) and Defense Advanced Research Projects Agency (DARPA) or the U.S. Government.}
\thanks{All procedures in animal experiments complied with the National Institutes of Health standards and were approved by the Animal Care and Use Committee of Rice University (Protocol\# IACUC-20-181).}
\thanks{All authors are with Rice University, Houston, TX 77005, USA. J. T. Robinson is also with Baylor College of Medicine, Houston, TX 77030, USA.}
}

%
%

\markboth{IEEE Journal of Solid-State Circuits}%
{Shell \MakeLowercase{\textit{et al.}}: Bare Demo of IEEEtran.cls for IEEE Journals}
%



\maketitle
\begin{abstract}
This paper presents a hardware platform including stimulating implants wirelessly powered and controlled by a shared transmitter for coordinated leadless multisite stimulation.
The adopted novel single-transmitter, multiple-implant structure can flexibly deploy stimuli, improve system efficiency, easily scale stimulating channel quantity and relieve efforts in device synchronization.
In the proposed system, a wireless link leveraging magnetoelectric effects is co-designed with a robust and efficient system-on-chip to enable reliable operation and individual programming of every implant. Each implant integrates a 0.8-mm$^2$
chip, a 6-mm$^2$ magnetoelectric film, and an energy storage capacitor within a 6.2-mm$^3$ size.
Magnetoelectric power transfer is capable of safely transmitting milliwatt power to devices placed several centimeters away from the transmitter coil, maintaining good efficiency with size constraints and tolerating 60-degree, 1.5-cm misalignment in angular and lateral movement. 
The SoC robustly operates with 2-V source amplitude variations that spans a 40-mm transmitter-implant distance change, realizes individual addressability through physical unclonable function IDs, and achieves 90\% efficiency for 1.5-to-3.5-V stimulation with fully programmable stimulation parameters.
\end{abstract}


%
\IEEEpeerreviewmaketitle

\section{Introduction}
%
%
%
%

\IEEEPARstart{M}{ultisite}  
biomedical stimulations that synchronously modulate activities of specific cells have shown exciting promise in clinical therapies. As two representative examples, multisite spinal cord stimulation shows the ability to restore patients motor function by stimulating different segments of the spine \cite{harkema_effect_2011, gerasimenko_initiation_2015}, 
and synchronized multisite cardiac pacing can  modulate contractions of different chambers to treat heart failures \cite{abraham_cardiac_2002, lyu_synchronized_2020, bereuter_leadless_2018}.
To reduce infection risks, surgery complexity, and restrictions in subject mobility, implantable biomedical stimulators need to have a miniaturized and untethered form factor. Moreover, for coordinated multisite stimulation, it is highly desirable that the system can flexibly deploy stimuli without leads, synchronize the operations of all the implants, and easily scale the number of stimulation channels.


\begin{figure}[t]
      \centering
      \includegraphics[width=0.95\linewidth]{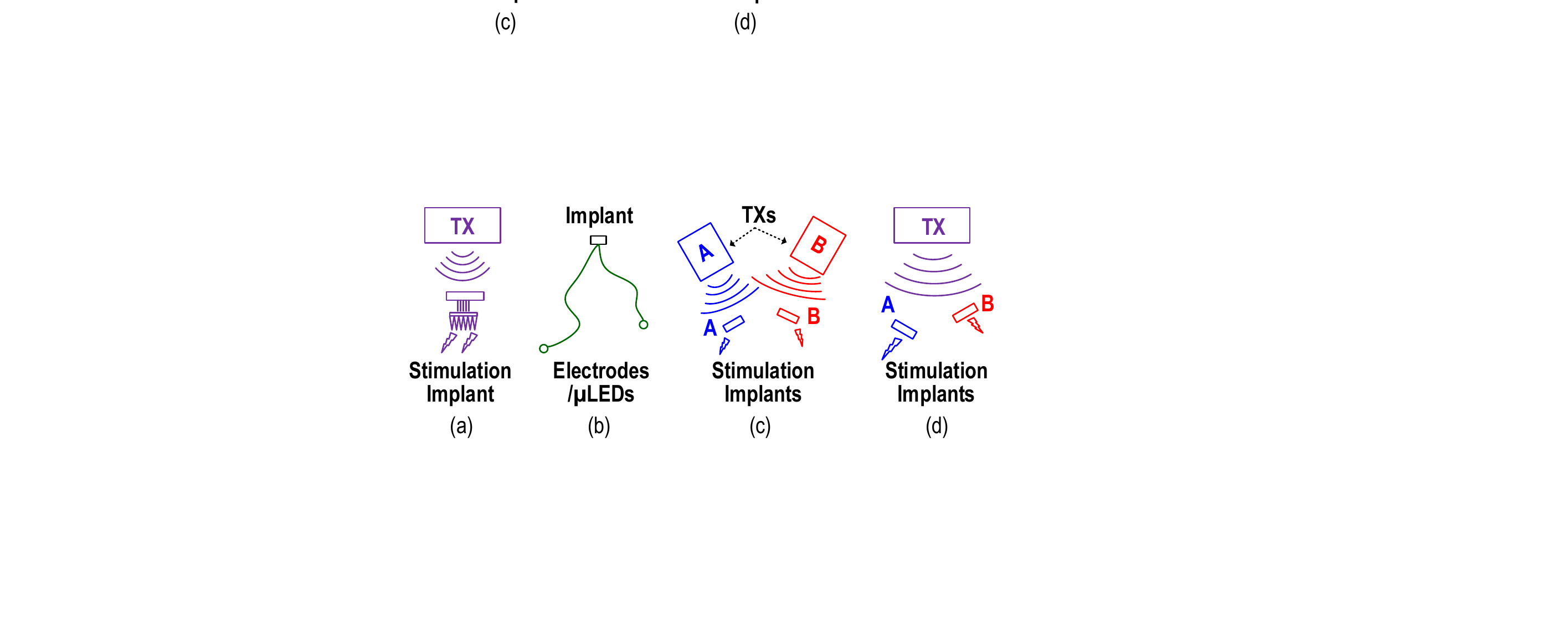}
      \caption{Various configurations of the multisite biomedical stimulation: (a) Implant with integrated electrode or \si{\micro}LED array, (b) Implant with wired electrodes or \si{\micro}LEDs, (c) TXs and implants pairs and (d) the proposed multiple implants powered and controlled by a single TX.}
      \label{S1_Comparison}
   \end{figure}

While significant progress has been made to develop bio-implants for multisite stimulation, there remain critical limitations to existing approaches.
Equipping an implant with electrode/LED arrays is a straightforward approach to add extra stimulation channels \cite{lo_176-channel_2016, jia_mm-sized_2018, jia_268_2020}, but the deployment of stimuli is limited by physical dimensions of the device, which are usually on the millimeter scale (Fig~\ref{S1_Comparison}~(a)).  
Another bottleneck is the fixed channel quantity of the electrode/LED array, which makes it infeasible to cheaply change the number of stimulating sites. 
While attaching leads to the implant can improve the flexibility of stimuli deployment \cite{gutruf_wireless_2019}, it may increase the risk of infection and interfere with the natural behavior of subjects (Fig~\ref{S1_Comparison}~(b)). 
Recently, a two-site heart pacing system was proposed with two independently powered and controlled implants to flexibly deploy stimuli without leads \cite{lyu_synchronized_2020}. However, the implants are inductively powered by two separate transmitters (TXs) through frequency multiplexing, as a result, they may face stricter electromagnetic exposure constraints for power transmission, more challenging device synchronization, and hence suffer limited capacity to support more implants (Fig~\ref{S1_Comparison}~(c)).

To circumvent these problems, we propose a coordinated multiple-site bio-stimulating system whose implants are wirelessly powered and controlled by a single TX, as illustrated in Fig~\ref{S1_Comparison}~(d). 
Since multiple wireless implants are employed, stimuli can be flexibly deployed without device size constraints.
Furthermore, the shared power and control link can significantly reduce difficulties in synchronizing all the implantable devices, improves overall power efficiency of the system and easily scale the number of stimulating sites.
In realizing the proposed technology, we need to tackle the following major challenges. First, the wireless power transfer mechanism must safely and efficiently deliver power to multiple implants deep inside the body. 
Second, body-movement-induced variations of TX-implant distance and misalignment are unavoidable in practice, especially when powering multisite implants using a single TX; they lead to varying input voltage and power for different implants. 
Lastly, every implant needs to be individually addressable and programmable with the shared transmitter for flexible and effective therapies. 

\begin{figure}[t]
      \centering
      \includegraphics[width=0.95\linewidth]{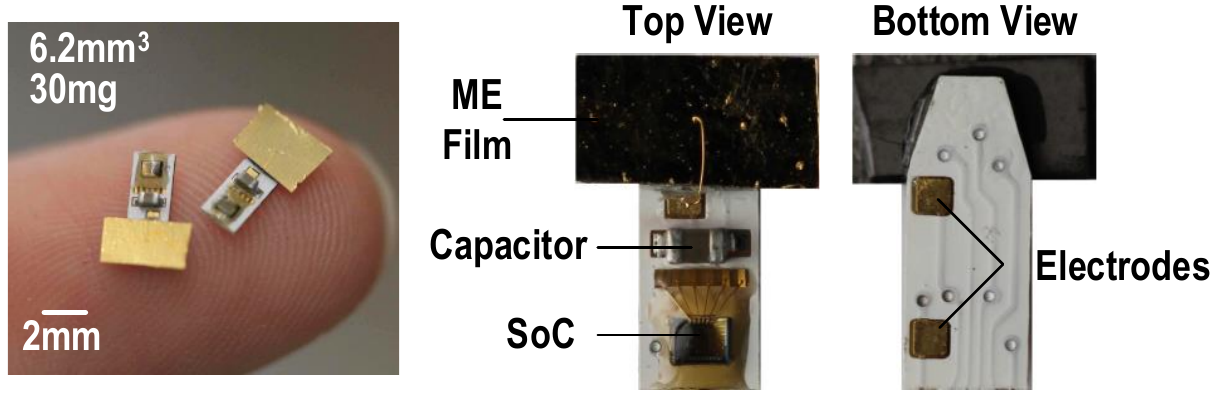}
      \caption{The proposed implants for coordinated multisite stimulation.}
      \label{S1_Implant}
   \end{figure}

To address these challenges, we developed a 6.2-mm$^3$, 30-mg implantable stimulator with a system-on-chip (SoC) that exploits magnetoelectric (ME) mechanisms for robust and efficient wireless powering, programming, and coordinated stimulation (Fig.~\ref{S1_Implant}). 
The presented stimulation implants feature: (1) reliable and synchronized operation under 2-V source variations, demonstrating a tolerance of 40-mm TX-implant distance variations; 
(2) robustness against up to 50-degree angular misalignment, 1.5-cm lateral misalignment with a 3-cm TX-implant separation;
(3) individual addressability and programmability enabled by on-chip physical unclonable function (PUF) IDs; (4) 90\% efficiency for stimulation with more than 1.5~V amplitude; and (5) fully programmable stimulation patterns covering 0.3-to-3.5-V amplitude, 0-to-1000-Hz frequency, 0.15-to-1.2-ms pulse width and 0-to-0.8-ms delay. 
These features make the proposed work suitable for clinical multisite spinal cord stimulation and multisite cardiac pacing, which usually require large stimulation amplitude ($\ge$ 0.5~V) \cite{harkema_effect_2011, lyu_synchronized_2020, bereuter_leadless_2018} and centimeter-scale implantation depth \cite{harrison_depth_1985,rahko_evaluation_2008} and stimulating site separation \cite{bereuter_leadless_2018, toossi_comparative_2021}.

   \begin{figure}[t]
      \centering
      \includegraphics[width=1\linewidth]{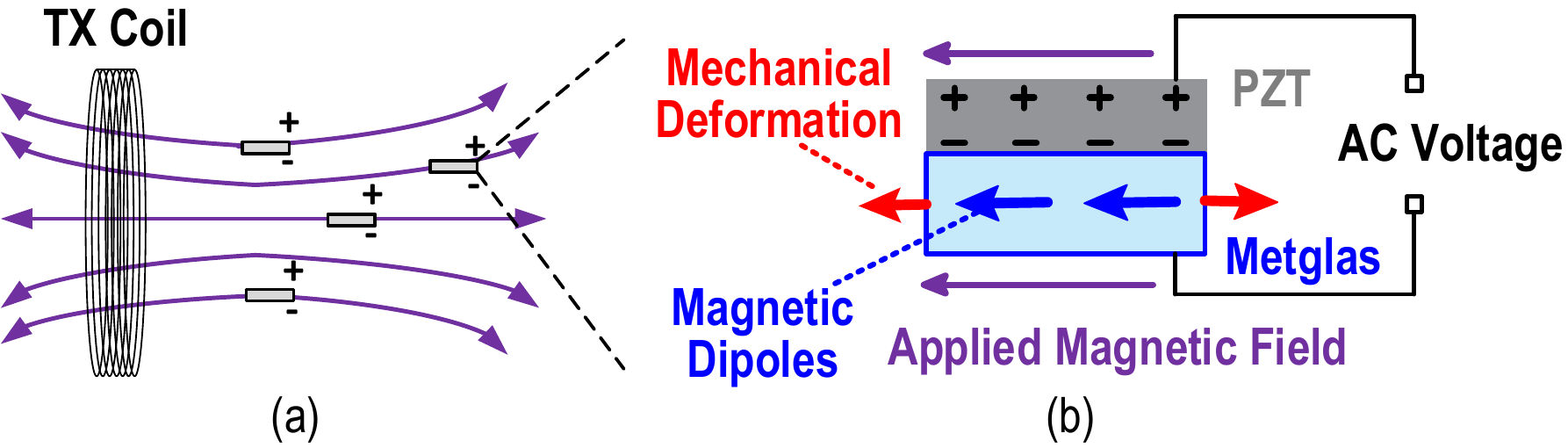}
      \caption{(a) Conceptual view of multiple ME films in a coil generated magnetic field and (b) operating principles of the PZT/Metglas-based ME transducer.}
      \label{S2_ME}
   \end{figure}

This paper is an extended version of \cite{yu_multisite_2021}, with more comprehensive qualitative and quantitative analysis of ME power transfer to multiple devices, detailed discussions on circuit implementations, and additional experimental results on the system robustness. The rest of the paper is organized as follows: Section II presents the design of the proposed system; Section III gives detailed implementations of the implant SoC; Section IV shows the experimental results, including functional validations, \textit{in-vitro} tests and \textit{in-vivo} experiments with animal models; Section V concludes this paper.

\section{System Design}
\label{sec:System}

The proposed multisite stimulation system is enabled by 
exploiting (1) an emerging wireless power transfer mechanism using magnetoelectric (ME) effects \cite{singer_magnetoelectric_2020, yu_magni_2020} to safely and efficiently deliver sufficient power for all the implants at different locations, and (2) a SoC that reliably and efficiently interfaces with the ME transducer to deliver consistent, synchronized, and individually programmed multisite stimulation, under significant motion-induced variations of TX-implant distance and misalignment. 

\begin{figure}[t]
      \centering
      \includegraphics[width=0.7\linewidth]{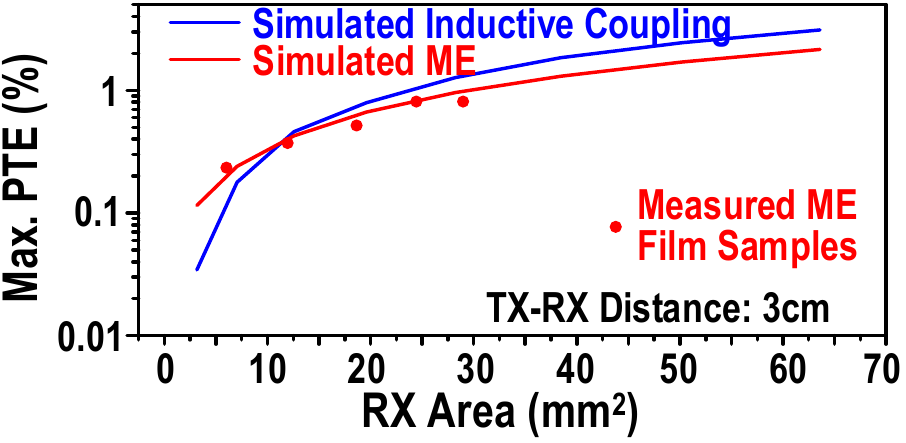}
      \caption{Efficiency comparison between the megnetoelectric and the inductive coupling with various RX size; the inductive coupling efficiency is simulated  and optimized with methods developed by \cite{ahn_optimal_2016}; measured efficiency of the ME samples are also added to the curve.}
      \label{S2_ME_Eff}
   \end{figure}
   
\begin{table}[t]
\caption{\textbf{Comparisons with various wireless power transfer for mm-sized RXs.}}
\label{table_PTE}
\centering
\setlength{\tabcolsep}{3.6pt}
\renewcommand{\arraystretch}{1.5}
\begin{tabular}{|p{45pt}|p{60pt}|p{40pt}|p{60pt}|}
\hline

\parbox[c][0.8cm]{45pt}{\centering
{Mechanism}} &
\parbox[c][0.8cm]{60pt}{\centering
{PTE at Depth}} &
\parbox[c][0.8cm]{40pt}{\centering
{$f_\mathrm{carrier}$}} &
\parbox[c][0.8cm]{60pt}{\centering
{RX Size}} 
\\\hline

\parbox[c][0.8cm]{45pt}{\centering
{ME \\ (This Work)}} &
\parbox[c][0.8cm]{60pt}{\centering
{1.67\% at 10 mm \\ 0.28\% at 30 mm}} &
\parbox[c][0.8cm]{40pt}{\centering
{0.33 MHz}} &
\parbox[c][0.8cm]{60pt}{\centering
{6 x 0.2 mm$^3$}} 
\\\hline

\multicolumn{1}{|c|}{Inductive \cite{zargham_fully_2015}}
& \multicolumn{1}{c|}{1.42\% at 10 mm}
& \multicolumn{1}{c|}{187 MHz}
& \multicolumn{1}{c|}{4.4 mm$^2$ (4-Turn)} 
\\\hline

\multicolumn{1}{|c|}{Inductive \cite{kim_144-mhz_2017}}
& \multicolumn{1}{c|}{2.04\% at 10 mm}
& \multicolumn{1}{c|}{144 MHz}
& \multicolumn{1}{c|}{8.6 mm$^2$ (2-Turn)} 
\\\hline

\multicolumn{1}{|c|}{Inductive \cite{kim_3_2019}}
& \multicolumn{1}{c|}{3.39\% at 10 mm}
& \multicolumn{1}{c|}{144 MHz}
& \multicolumn{1}{c|}{8.7 mm$^2$ (3-Turn)} 
\\\hline

\multicolumn{1}{|c|}{Inductive \cite{jia_dual-band_2019}}
& \multicolumn{1}{c|}{4.1\% at 5 mm}
& \multicolumn{1}{c|}{60 MHz}
& \multicolumn{1}{c|}{4.9 mm$^2$ (6-Turn)} 
\\\hline


\multicolumn{1}{|c|}{Ultrasound \cite{piech_wireless_2020}}
& \multicolumn{1}{c|}{0.06\% at 18 mm}
& \multicolumn{1}{c|}{1.85 MHz}
& \multicolumn{1}{c|}{0.52 mm$^3$} 
\\\hline

\end{tabular}
\label{table_PTE}
\end{table}

\subsection{Magnetoelectric Power Transfer to Multiple Implants}
\label{subsec:System_ME}

Magnetoelectric (ME) thin-film transducers convert low-frequency alternating magnetic fields into electrical energy via mechanical coupling between the magnetostrictive (Metglas) and piezoelectric (PZT) laminates, as shown in Fig.~\ref{S2_ME} \cite{dong_longitudinal_2003, singer_magnetoelectric_2020}. 
Because of the mechanical coupling, a 2x3-mm$^2$ ME film resonates at merely 330~kHz, while generating a voltage greater than 7~Vpp when the magnetic field strength is larger than 5 Oe, with a purely resistive source impedance around 800-\si{\Omega}. These properties make ME a promising candidate for efficient and safe wireless power transfer to miniaturized bio implants with low tissue absorption and reflection~\cite{yu_magni_2020,yu34.3,Singer_wireless}. This work quantitatively studies and exploits the efficiency, robustness, power limits, and scalability of the ME-based wireless power transfer to realize the proposed multisite stimulation system.

   \begin{figure}[t]
      \centering
      \includegraphics[width=1\linewidth]{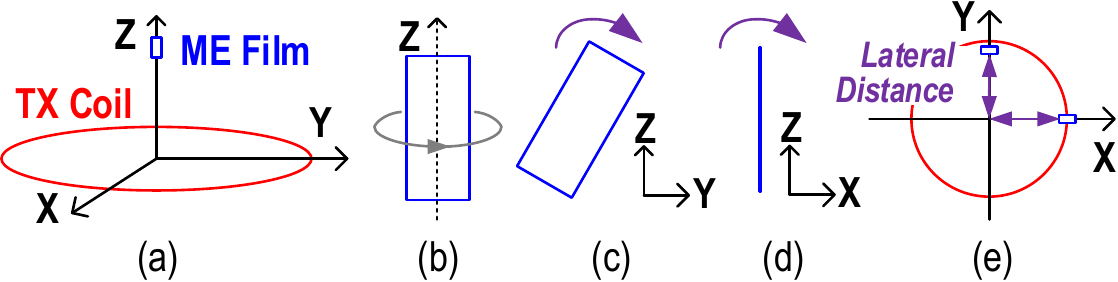}
      \caption{Illustrations for (a) the ideal alignment; the angular misalignment with rotations (b) along the Z-axis, (c) in the YZ-plane, (d) in the XZ-plane; and (e) the lateral misalignment with movement along the X-axis and the Y-axis in the ME power transfer.}
      \label{S2_Mis}
   \end{figure}
   
   \begin{figure}[t]
      \centering
      \includegraphics[width=1\linewidth]{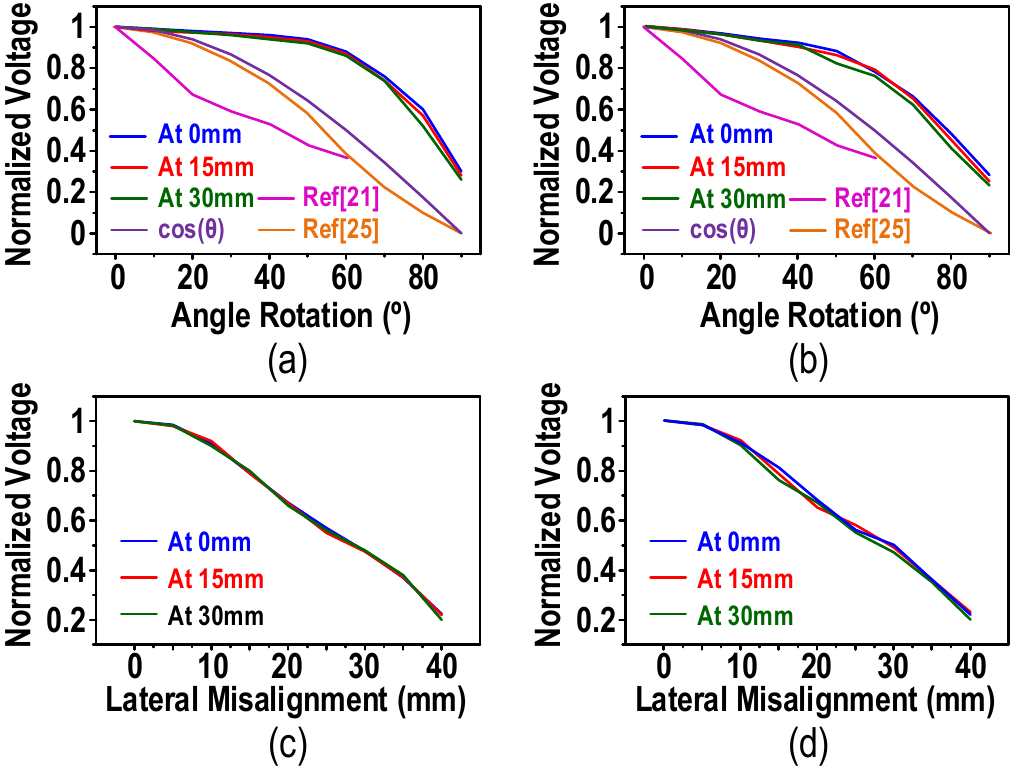}
      \caption{Measured normalized ME voltage with (a) angular rotation in the XZ-plane, (b) angular rotation in the YZ-plane, (c) lateral movement along the x-axis and (d) lateral movement along the y-axis at three different distances; angular misalignment sensitivity of ultrasonic \cite{piech_wireless_2020} and inductive coupling \cite{burton_wireless_2020} power transfer are included for comparisons.}
      \label{S2_Mis_Plot}
   \end{figure}

\subsubsection{High Power Transfer Efficiency to Miniaturized Receivers}
The ME voltage coefficient (unit: V/Oe) determining the ME induced voltage magnitude for a given magnetic field strength is independent of film width and length; shrinking film area only changes the resonant frequency  and the intrinsic impedance, which are inversely proportional to the film length and width, respectively \cite{zhou_uniform_2014, yu_magni_2020}. 
As a result, the ME films have great potential to maintain good efficiency with miniaturized dimensions. 
In comparison with the most relevant low-frequency magnetic-field-based inductive coupling, the ME power transfer efficiency (PTE) displays a slower declining rate with area reduction and outperforms the inductive coupling when the area is below 10 mm$^2$ (Fig~\ref{S2_ME_Eff}). 
The PTE is defined as the ratio of power received at the receivers (RX coils or ME transducers) through power consumed by the TX coil.
The ME efficiency reported here is simulated in COMSOL, and several measurement results of ME film samples are also added. The inductive coupling curve is simulated and optimized in HFSS using methods introduced by \cite{ahn_optimal_2016}, where the maximum efficiencies at the optimal operating frequencies are utilized for fair comparisons.
Furthermore, we compare the ME with various wireless power transfer mechanisms exploited by recent work for mm-sized RXs in Table \ref{table_PTE} \cite{zargham_fully_2015,kim_144-mhz_2017,kim_3_2019,jia_dual-band_2019, piech_wireless_2020}.
The ME shows comparable or better PTE at a cm-scale implantation depth but operates at a much lower frequency than the inductively powered work, resulting in alleviated EM exposure constraints.
This advantage will be further discussed in the following part.







\subsubsection{Low Sensitivity to Misalignment}
Guaranteeing an ideal alignment is difficult, especially in the single-transmitter, multiple-receiver scenarios. Because of this issue, stability with misalignment becomes an essential consideration in establishing the wireless link in this work.
Near-field inductive coupling is known to be sensitive to perturbations in orientation when the coils are small \cite{fotopoulou_wireless_2011, burton_wireless_2020}. Theoretically, in a uniform magnetic field, RX coil voltage is proportional to cos(\si{\theta}), where \si{\theta} is RX's rotating angle \cite{burton_wireless_2020}.
The ME voltage is proportional to the magnetic field strength in the direction of the transducer’s long axis.
Due to the ME transducer's physical characteristics, angular misalignment can happen when the film rotates along the Z-axis, in the YZ-plane ,or in the XZ-plane, (Fig~\ref{S2_Mis}~(b), (c), and (d)),
while the first case will not affect the voltage owing to the symmetry of the magnetic field. In addition to these, misalignment can happen when the ME films laterally move from the TX coil center (Fig~\ref{S2_Mis}~(e)). 

Because of the high permeability of Metglas, the ME laminates own significant magnetic flux concentration effect, which can change the direction of the magnetic flux line inside the materials and make the angle between the transducer’s long axis and the flux line smaller than \si{\theta}~\cite{fang_enhancing_2009, yu_magni_2020}.
This feature can alleviate the influences of angle rotations and lead to a voltage higher than $V_\mathrm{0}\times$cos(\si{\theta}), where $V_\mathrm{0}$ is the voltage with ideal alignment.
Here, for a comprehensive study, we measure the ME voltage when the film rotates in the XZ- or YZ-plane at different vertical distances from the TX coil, as given in Fig. \ref{S2_Mis_Plot} (a) and (b), in which the function cos(\si{\theta}) curve is added for reference as well. In all conditions, the ME demonstrates a less-than-20\% voltage reduction if angle rotation is smaller than 60 degrees, and much slower decline rate than cos(\si{\theta}).
It also shows less sensitivity to angular misalignment when compared to state-of-the-art ultrasonic~\cite{piech_wireless_2020} and inductive coupling~\cite{burton_wireless_2020} power transfer.
Effects of lateral movements of the ME films are studied as well, which show \si{\sim} 20\% reduction in voltage for a lateral distance of 1.5 cm (Fig. \ref{S2_Mis_Plot}~(c) and (d)). No substantial difference is observed in the measurement curves when moving along the x-axis and y-axis.
We will further appraise the misalignment sensitivity of the proposed devices in Section~\ref{subsec:T_Ele}. 

\subsubsection{High Power Transmission under Safety Limits}
In this work, the ME films have acoustic resonant frequencies near 330~kHz. 
The low carrier frequency leads to low tissue absorption, allowing a more than ten times stronger magnetic field to penetrate the body without violating the safety limits \cite{noauthor_ieee_nodate} (in comparison with a magnetic field at 13.56 MHz).
As a result, order-of-magnitude higher power can be safely transmitted deep inside the body through ME compared to the 13.56-MHz near-field inductive coupling.
At a 3-cm distance, the ME film can receive 3.8-mW peak power without safety issues, while the 13.56-MHz inductive coil at 3 cm obtains 0.4 mW (calculation based on simulations of the maximum allowed magnetic field strength and power transfer efficiency).

\begin{figure}[t]
      \centering
      \includegraphics[width=0.95\linewidth]{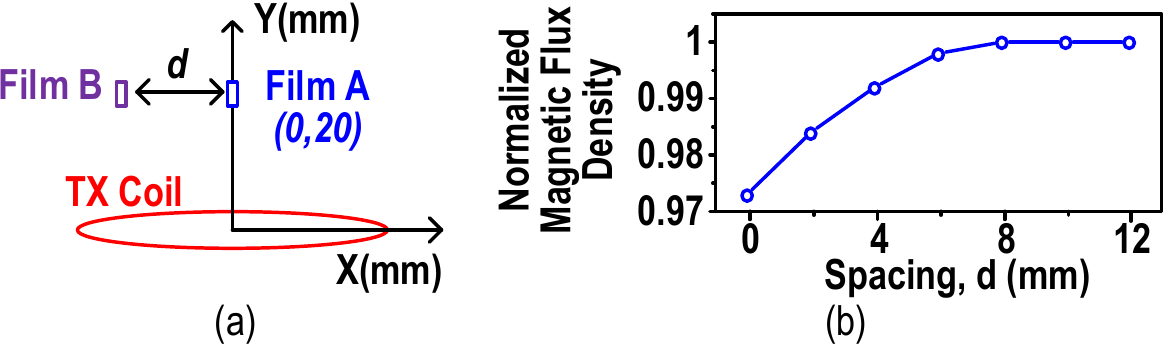}
      \caption{(a) Illustration for simulation to evaluate the limitations on ME device distance and (b) simulated magnetic flux density of film A versus RX separation.}
      \label{S2_Intra}
   \end{figure}

\subsubsection{Good Scalability}

The ME transducers generate voltage relying on the magnetic field strength and will not affect the TX's working conditions.  
As a result, including more devices in the system does not require increasing the power consumption in TX.
Thus, the maximum number of ME devices mainly depends on the dimensions of the region where devices can reliably work, implant size, and the minimum distance between implants.
Thanks to the strong penetration and the lateral misalignment robustness of the ME link and the mm-scale miniaturization of the RXs, the proposed system simultaneously achieves a large operating region and a small implant size, bringing benefits to include more devices.
Considering the ME laminates can concentrate the magnetic flux, some interactions between the closely placed films may exist, which may influence the flux density inside the films and limit the minimum implant distance. 
To evaluate this, we conduct a COMSOL simulation, in which we move film B towards film A at (0, 2 cm), and simulate the average magnetic flux density inside film A (Fig.\ref{S2_Intra}~(a)).
While a decrease in the magnetic flux density of film A is observed when the spacing is smaller than 8 mm, it is less than 3\% and should not affect the ME transducer’s regular operation (Fig.\ref{S2_Intra}~(b)).
This result is consistent with our experimental observations that no noticeable input voltage changes happen when placing the ME devices close together.
Therefore, the proposed technology has favorable scalability properties for supporting multiple implants.

\subsection{Overview of the Implant SoC}
\label{subsec:System_Implant}

\begin{figure}[t]
      \centering
      \includegraphics[width=1\linewidth]{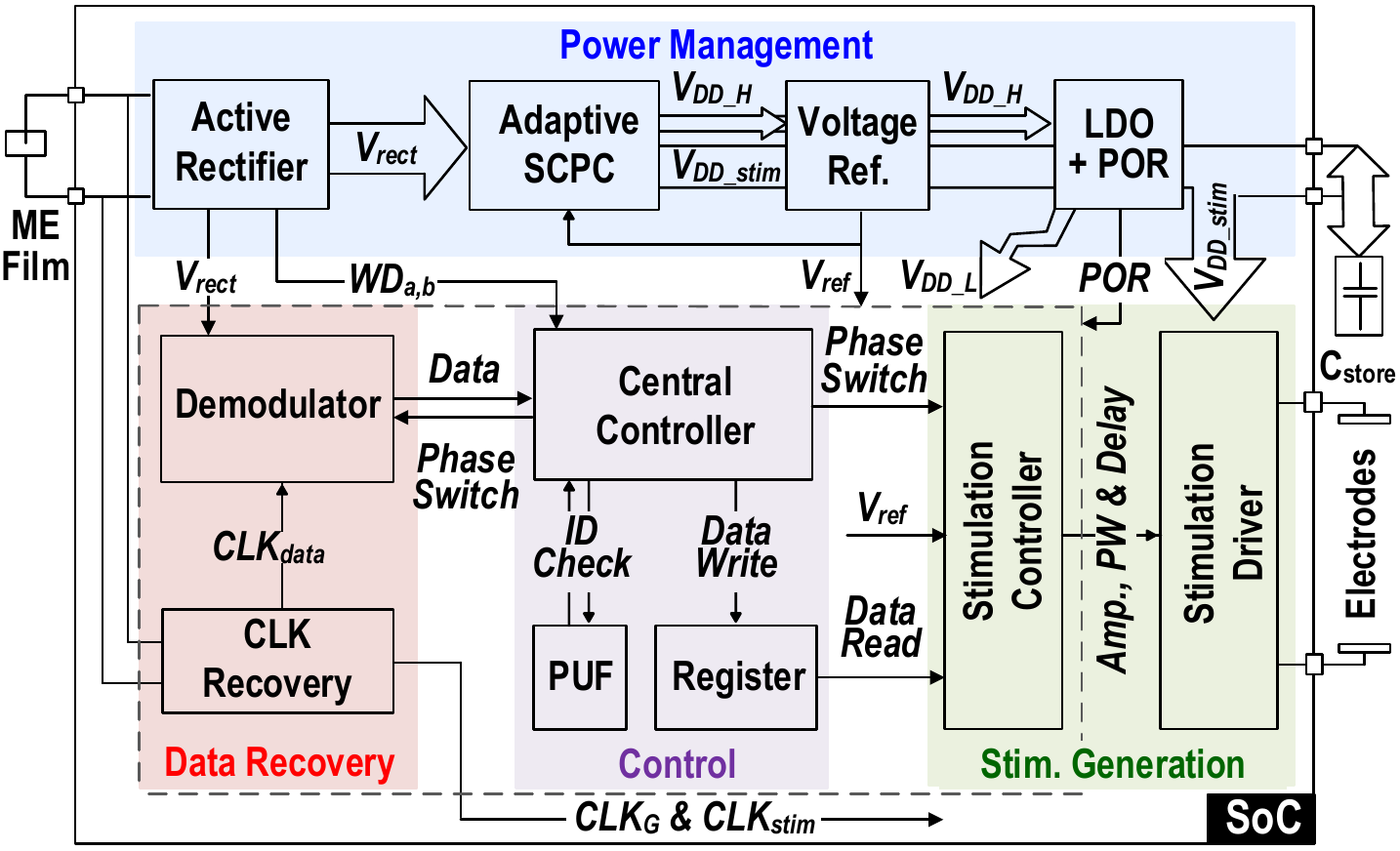}
      \caption{Block diagram of the implant SoC.}
      \label{S2_SoC}
   \end{figure}

The SoC interfaces with the ME transducer to harvest energy through ME effects. It consists of power management, control, data recovery and stimulation generation modules to allow each implant to receive data from the external TX and generate programmable stimulation.

In the power management, the ME induced voltage is rectified to $V_\mathrm{rect}$ and then converted by an adaptive switched-capacitor power converter (SCPC). The SCPC generates a proper voltage and buffers energy on the off-chip capacitor for stimulation and provides a high-voltage supply $V_\mathrm{DD\_H}$ for the reference generator and the low-dropout regulator (LDO). A 1-V supply $V_\mathrm{DD\_L}$ is generated by LDO for the control and data recovery circuitry and a power-on-reset (POR) signal is triggered when $V_\mathrm{DD\_L}$ stabilizes. Supply-invariant reference voltages for the entire system are generated by a ultra-low-power reference circuit with a native NMOS and stacked diode-connected PMOS transistors \cite{seok_portable_2012}.

The downlink data is modulated through the amplitude shift keying (ASK) and will be detected and decoded in data transmission phase by the demodulation circuit, whose input is the rectified voltage. Considering the settling time of the ME transducers, 64 resonance cycles are assigned to transmit one bit for reliable downlink data transfer, resulting in a data rate of 5.16 kbps.
The clock recovery circuits extract the global clock $CLK_\mathrm{G}$ at 330 kHz from the film transducer and provide timing references at desired frequencies for other functionalities (10.3 kHz for data recovery and 82.5 kHz for stimulation). 
A central controller is in charge of switching the operating phases of the SoC. It also checks the recovered data with the device ID to determine whether to update the data stored in the register file.
The accepted data will set the stimulation recipe by programming the stimulation controller, which contains a voltage digital-to-analog converter and a finite-state-machine to provide fully changeable patterns, covering amplitude, pulse width and delay. 
Desired stimuli is driven by the stimulation driver powered by $V_\mathrm{DD\_Stim}$ and delivered to the target tissue through the on-board electrodes.
\section{Implant System-on-Chip Implementation}
\label{sec:SoC}

To achieve the desired coordinated multisite stimulation, we implement the implant SoC with technical emphasis on adaptively and efficiently managing power for low-loss stimulation, operating with source-variation robustness and synchronization among all the implants, and individually addressing and programming each device. 

\subsection{Adaptive Power Conversion and Highly-Efficient Stimulation Generation}
\label{subsec:SoC_Power}

\begin{figure}[t]
      \centering
      \includegraphics[width=0.8\linewidth]{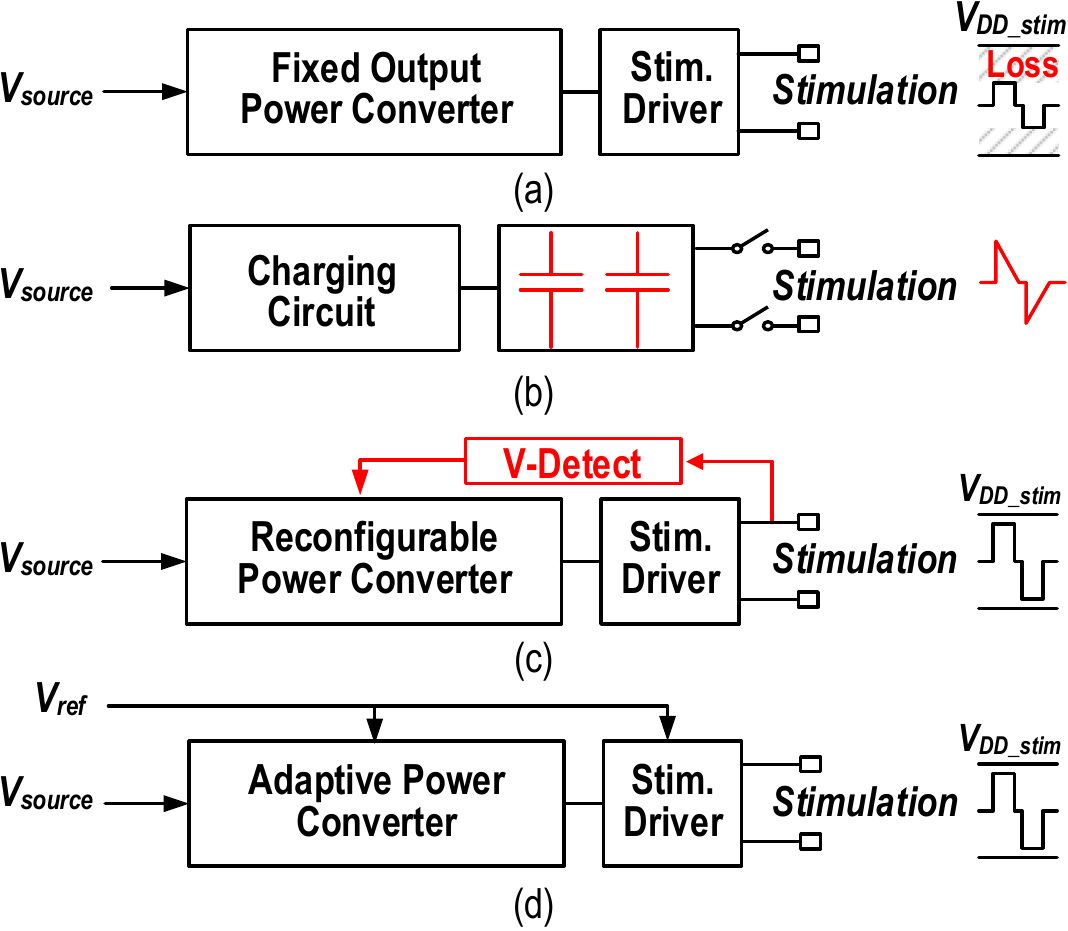}
      \caption{Existing power recovery and stimulation schemes: (a) power conversion providing fixed stimulation supply, (b) switched-capacitor based stimulation , and (c) closed-loop stimulation supply voltage control; (d) the proposed low-overhead adaptive power conversion for high-efficiency stimulation.}
      \label{S3_Comparison}
   \end{figure}
   
   \begin{figure}[t]
      \centering
      \includegraphics[width=0.95\linewidth]{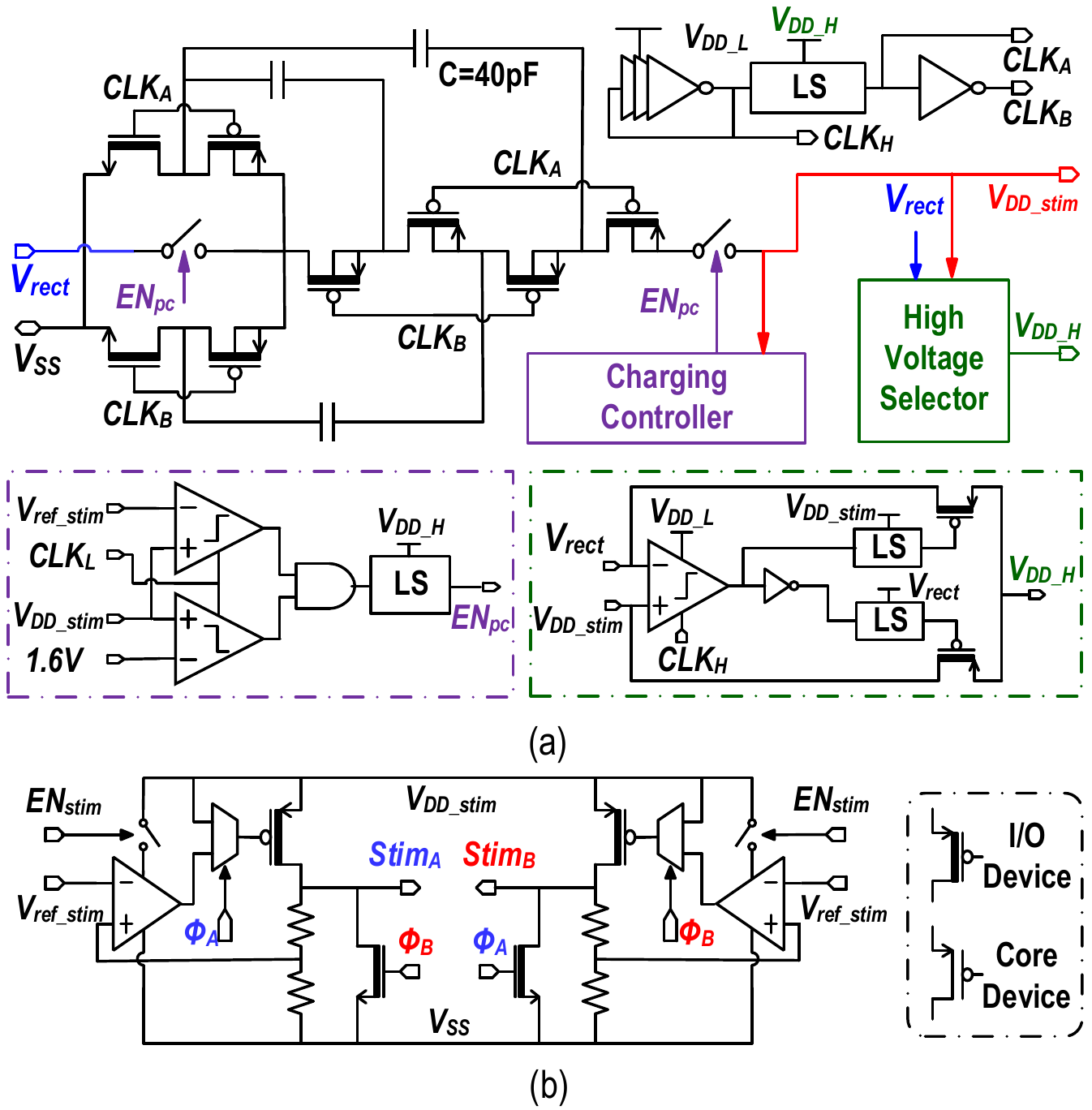}
      \caption{Implementations of (a) the adaptive switched-capacitor power converter and (b) the stimulation driver.}
      \label{S3_SCPC}
   \end{figure}

Power recovery and stimulation generation should be insensitive to source amplitude variations, especially when using multisite implants due to the high degree of variability of distance and misalignment with the TX.
Regulation of stimulation pulses is also desired to precisely control charge deposition into the tissue for safety. Furthermore, high efficiency in generating stimulation can help to alleviate heat generation of the devices. 

Simply generating a high enough supply may ensure sufficient stimulation power even with source variations but will suffer from huge power loss and thus heat dissipation (Fig.~\ref{S3_Comparison}~(a)) \cite{piech_wireless_2020}. 
Alternatively, unregulated voltage stimulation by directly driving electrodes with charged capacitors is efficient (Fig.~\ref{S3_Comparison}~(b)) \cite{lee_power-efficient_2015}. However, it sacrifices precise control of stimulation pulse width and waveform, which are important for stimulation effectiveness and safety \cite{gorman_effect_1983, merrill_electrical_2005}, and the capacitor bank may occupy a large area.
Previous work demonstrated regulating the stimulation with low power loss through real-time electrode voltage sensing \cite{lee_power-efficient_2013}. However, complex implementation and high power consumption might be added by the feedback and reconfiguration loop (Fig.~\ref{S3_Comparison}~(c)). 
Here, we propose a design to adaptively recover power to improve stimulation efficiency, where the same reference controls the generation of stimulation supply and amplitude, as shown in Fig.~\ref{S3_Comparison}~(d). The supply voltage is set to be 10\% higher than the stimulation amplitude, which ensures a 90\% efficiency with good regulation. The voltage reference can be simply programmed by the downlink data with low overhead. 

The adaptive power converter employs a switched-capacitor topology with a core of 4X charge pump, which ensures sufficient supply for stimulation even with low input voltage (Fig.~\ref{S3_SCPC}~(a)).
A ring oscillator generates a 900-kHz clock for the power converter. 
Regulation of the stimulation supply is realized by disconnecting the energy storage capacitor $C_\mathrm{store}$ from the power converter. Considering the charging and discharging of $C_\mathrm{store}$ is slow due to the large capacitance, low-speed clock $CLK_\mathrm{L}$ is utilized in the charging controller for low-power design.
The always-on voltage selector connects the output to the higher one of the rectified voltage $V_\mathrm{rect}$ and the stimulation supply $V_\mathrm{DD\_stim}$. It not only enables cold startup using $V_\mathrm{rect}$ but also guarantees that the system has a sufficient voltage source, especially when the ME voltage amplitude changes for ASK modulation. 

The stimulation driver adopts a LDO-based structure with high-speed amplifiers, as shown in Fig.~\ref{S3_SCPC}~(b). Those amplifiers are only powered in stimulation phase, consuming negligible power in comparison with stimulation. Considering that the amplifiers require proper supply to maintain good performance and device heating is not a substantial problem in low-power stimulation, the minimum supply of the stimulation driver is set as 1.5~V. As a result, stimulation holds less than 10\% power loss in 1.5-to-3.5-V voltage range.
The driver can work in either monophasic or biphasic mode programmed by the downlink data. While monophasic pulses may lead to a lower threshold and shorter latency, the biphasic stimulation can balance charge and prevent undesired electrochemical reactions on electrodes \cite{rubinstein_analysis_2001}. After each stimulus, the electrodes are shorted to remove the residual charge for stimulation safety.

\begin{figure}[t]
      \centering
      \includegraphics[width=0.9\linewidth]{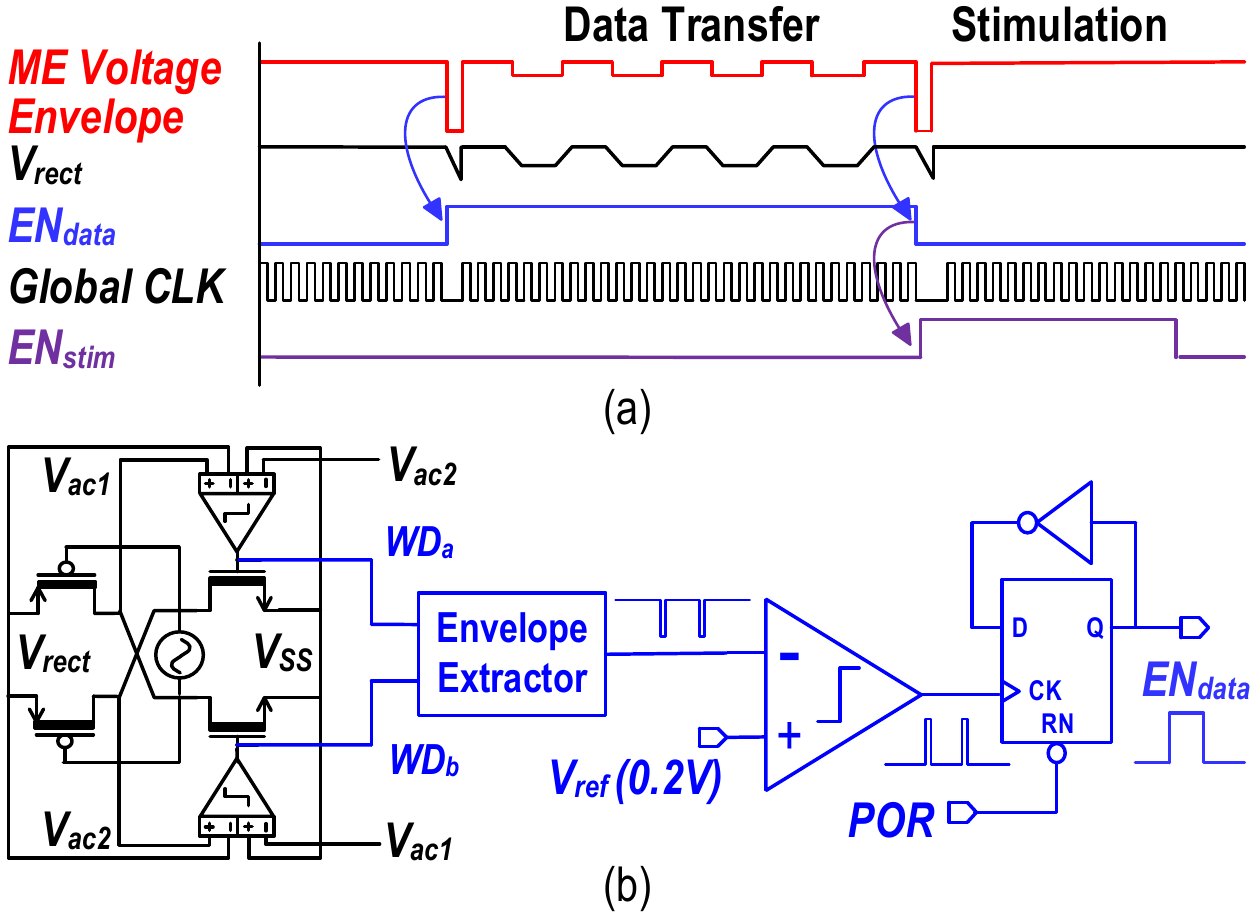}
      \caption{(a) Timing diagram of the system operation and (b) circuit implementation of operating phase controller.}
      \label{S3_Control}
   \end{figure}
   
\begin{figure}[t]
      \centering
      \includegraphics[width=0.95\linewidth]{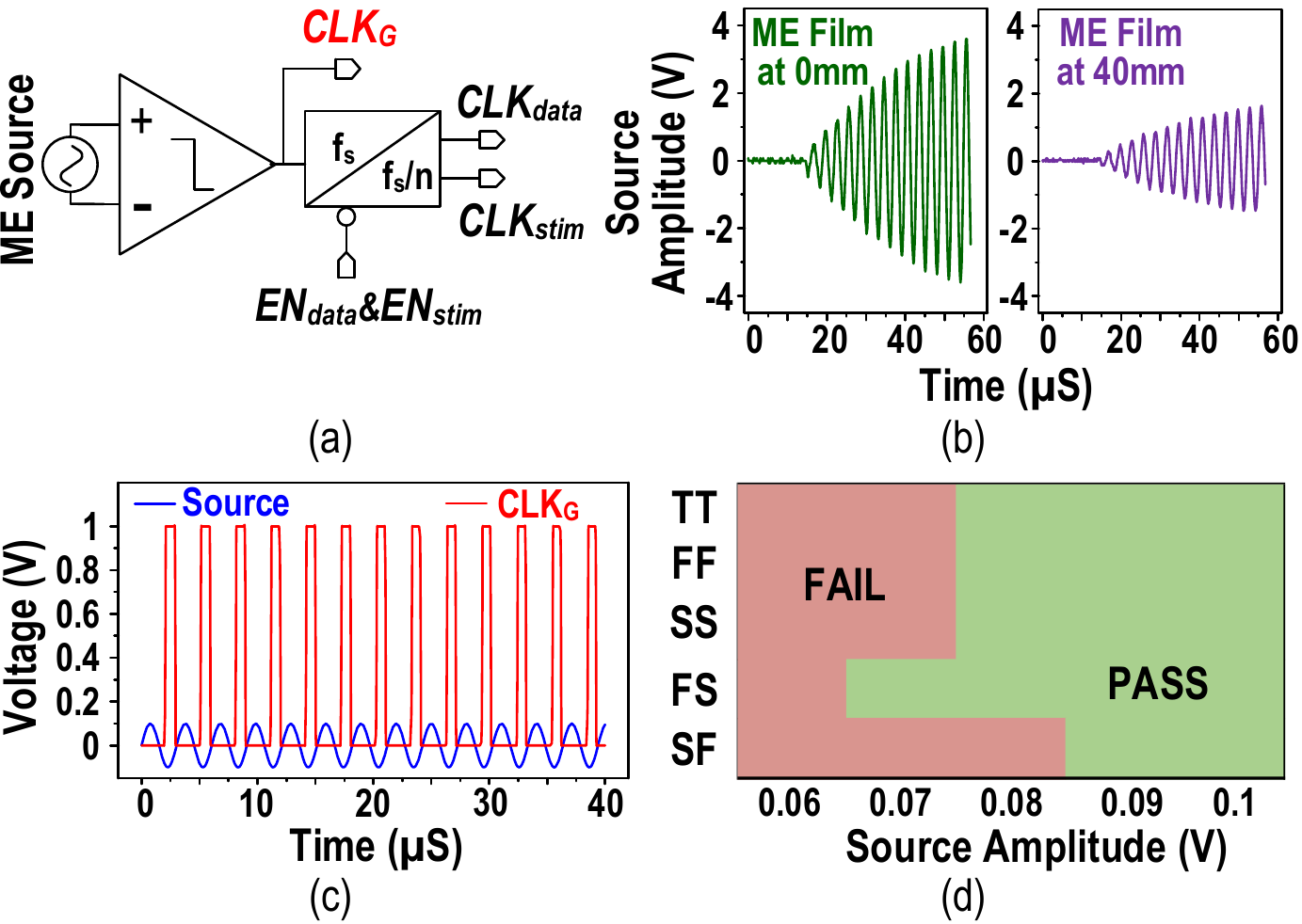}
      \caption{(a) Implementation of the clock recovery circuit, (b) measured ME voltage of two films when the magnetic field turns on, (c) simulation of clock recovery with a 100-mV source and (d) shmoo plot of simulated clock with varying source amplitude at five process corners.}
      \label{S3_CLK}
   \end{figure}

\subsection{Synchronized Operation with Robustness}
\label{subsec:SoC_Syn}

To realize synchronization, we adopt a strategy that the operation of all the implants are fully controlled by a shared TX. 
The command to switch the implant's operating phases is represented as the absence of the magnetic field. This absence can be a 100-\si{\micro}s short notch.
In one operating cycle, the first notch signifies the beginning of data transmission, and the second one triggers the generation of the stimulation (Fig.~\ref{S3_Control}~(a)). This notch-based control scheme is robust with TX-implant distance changes and misalignment, since it disregards the received voltage of each device, ensuring all the implants to work in the same phase.
To detect these notches, outputs of the 4-input comparators in the active rectifier \cite{lam_integrated_2006} are reused as watchdog signals $WD_\mathrm{a}$ and $WD_\mathrm{b}$, as shown in Fig.~\ref{S3_Control}~(b). 
The comparator outputs track the envelope of $V_\mathrm{rect}$ during the charging and become zero if the the magnetic field is absent, in which case the ME induced voltage disappears and the rectifier stops charging the load.

Meanwhile, the global clock signals of all implants are recovered from the same source to provide synchronized process-invariant timing references for data sampling and stimulation, as illustrated in Fig.~\ref{S3_CLK}~(a).
Like inductive coils, the ME films also have a ring-up behavior that requires nearly 15 response cycles to settle to the desired voltage when the magnetic field resumes. 
Consequently, the source amplitude might be small during ramping up, especially for those films far from the external TX, introducing challenges in recovering clocks for all the implants at the same time.
Fig.~\ref{S3_CLK}~(b) shows when turning the magnetic field on, the voltage of a film that is 4-cm away from the TX can be as small as 160 mV. 

\begin{figure}[t]
      \centering
      \includegraphics[width=0.95\linewidth]{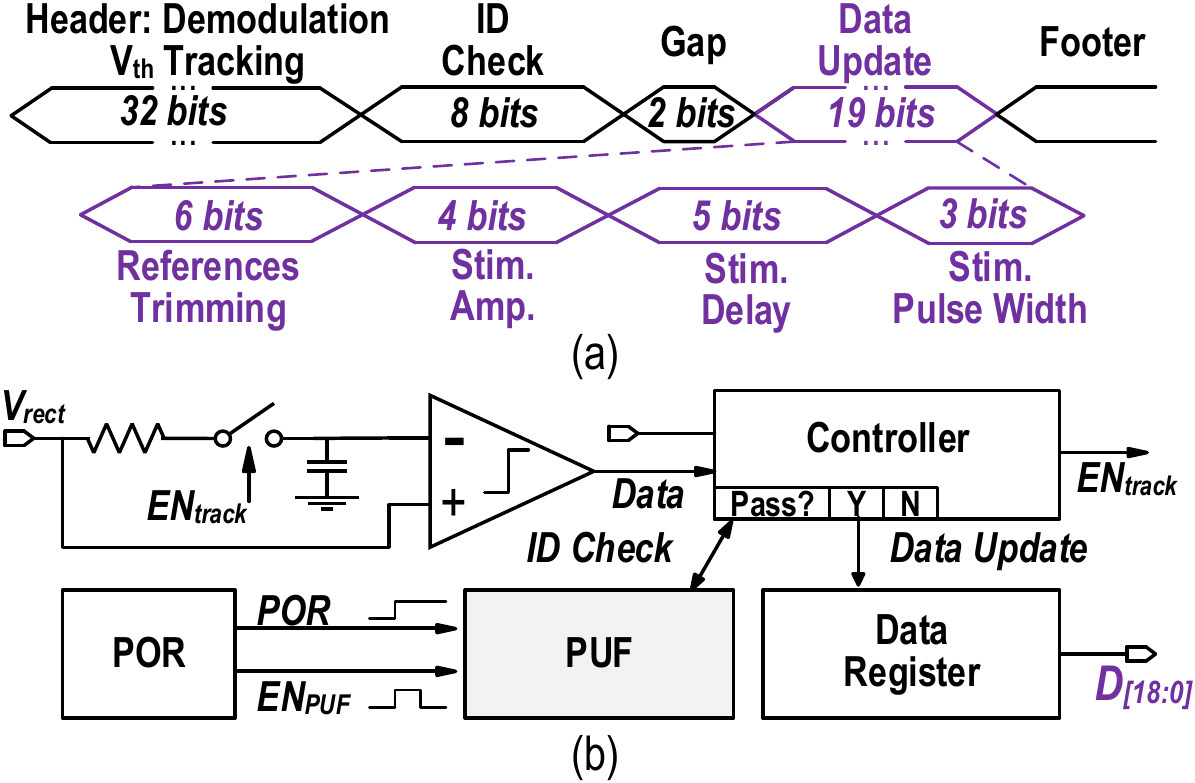}
      \caption{(a) Diagram of the data packet containing device ID and (b) realization of individual programming through ID check with the PUF IDs. }
      \label{S3_Data}
   \end{figure}

\begin{figure}[t]
      \centering
      \includegraphics[width=0.9\linewidth]{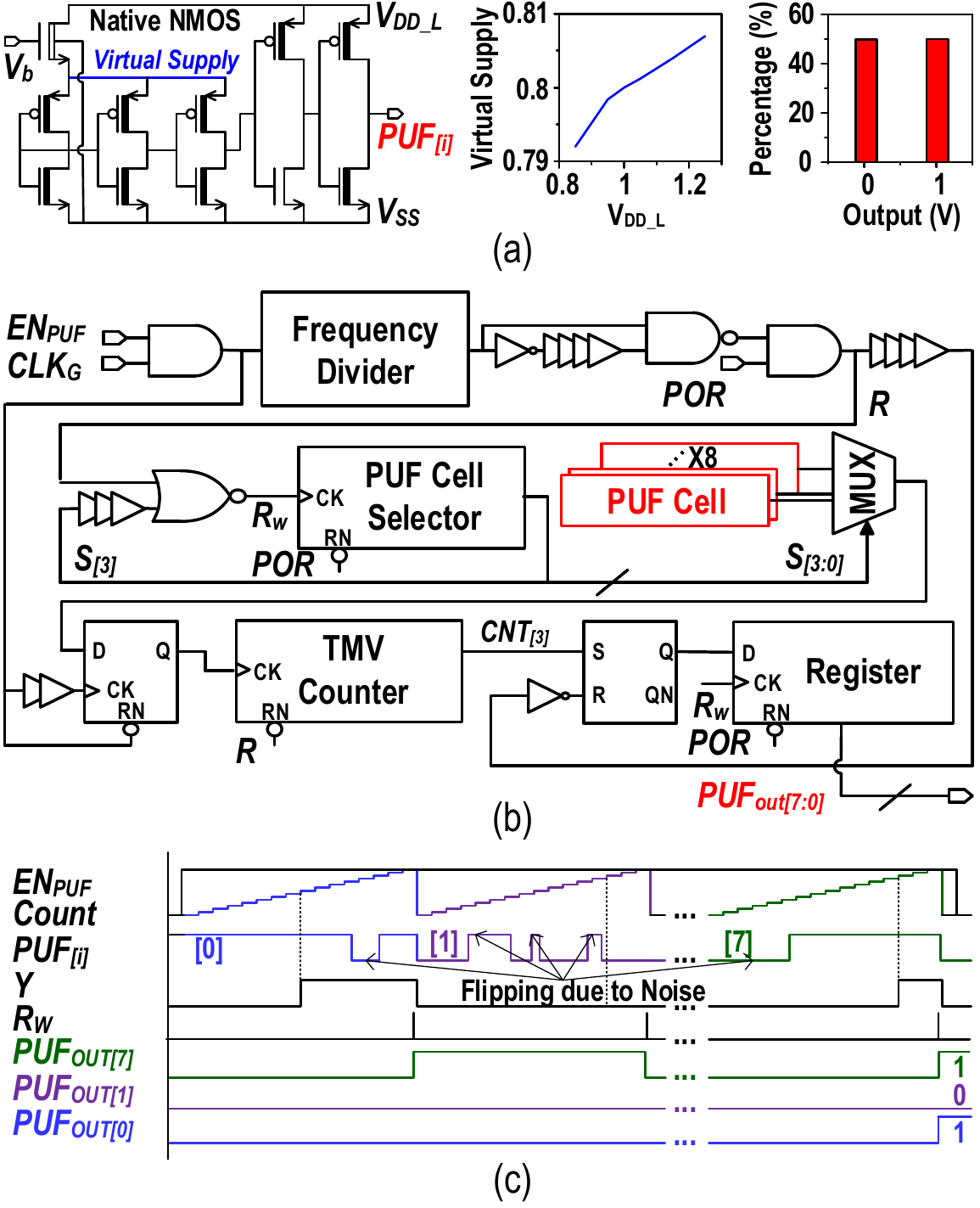}
      \caption{(a) Schematic of the PUF cell, simulated self-regulated virtual supply of the PUF cell versus the digital-domain supply $V_\mathrm{DD\_L}$, and the percentage of its simulated output based on 5000 iterations, (b) implementation of the 8-bit PUF with temporal majority voting and (c) timing diagram of the PUF's operation.}
      \label{S3_PUF}
   \end{figure}

In order to avoid missing any clock cycles, the comparator here is designed with PMOS as input transistors, achieving a high sensitivity when the input amplitude is low. 
As shown in Fig.~\ref{S3_CLK}~(c), it successfully detects the voltage difference of a 100-mV source and generates pulses at 330~kHz as expected.
The pulses may be produced at a different time due to the dependence on source amplitude, leading to potential phase variations in the clocks of implants at various positions.
The worst case is that one implant generates its first clock pulse when the phase of the ME sinusoidal output is close to 0 degrees while another implant starts recovering the clock when ME voltage reaches the peak of a single cycle.
However, even considering the worst case, the timing difference between these two clocks is merely 0.75~\si{\micro}s, i.e., a quarter of the carrier period.
It is negligible compared to the pulse width of a single data bit and the minimum stimulation pulse duration, which are 194~\si{\micro}s and 50~\si{\micro}s.
The simulation results in Fig.~\ref{S3_CLK}~(d) show that the clock can be correctly extracted with a source amplitude as low as 90~mV in five process corners. 
To further ensure the clocks for data recovery and stimulation generation are synchronized among all implants, the frequency divider is reset at the beginning of data transmission and stimulation phases.

\subsection{PUF Enabled Individual Programming}
\label{subsec:SoC_PUF}

Individually programming each implant is critical for effective and flexible stimulating therapies. 
To achieve this with a shared TX, we design a data packet containing 8-bit ID for device addressing (Fig.~\ref{S3_Data}~(a)). 
The data packet also has a preamble for autonomous demodulation threshold calibration to enable successful data recovery with different ME voltages and a 19-bit data payload to calibrate the voltage reference affected by the process variations and program the stimulating settings.
The demodulation threshold is extracted by a low-pass filter and a track-and-hold circuit.
Once the data is recovered, the update controller will check the received ID against the on-chip ID to decide whether to accept the new payload data (Fig.~\ref{S3_Data}~(b)). 

\begin{figure}[t]
      \centering
      \includegraphics[width=1\linewidth]{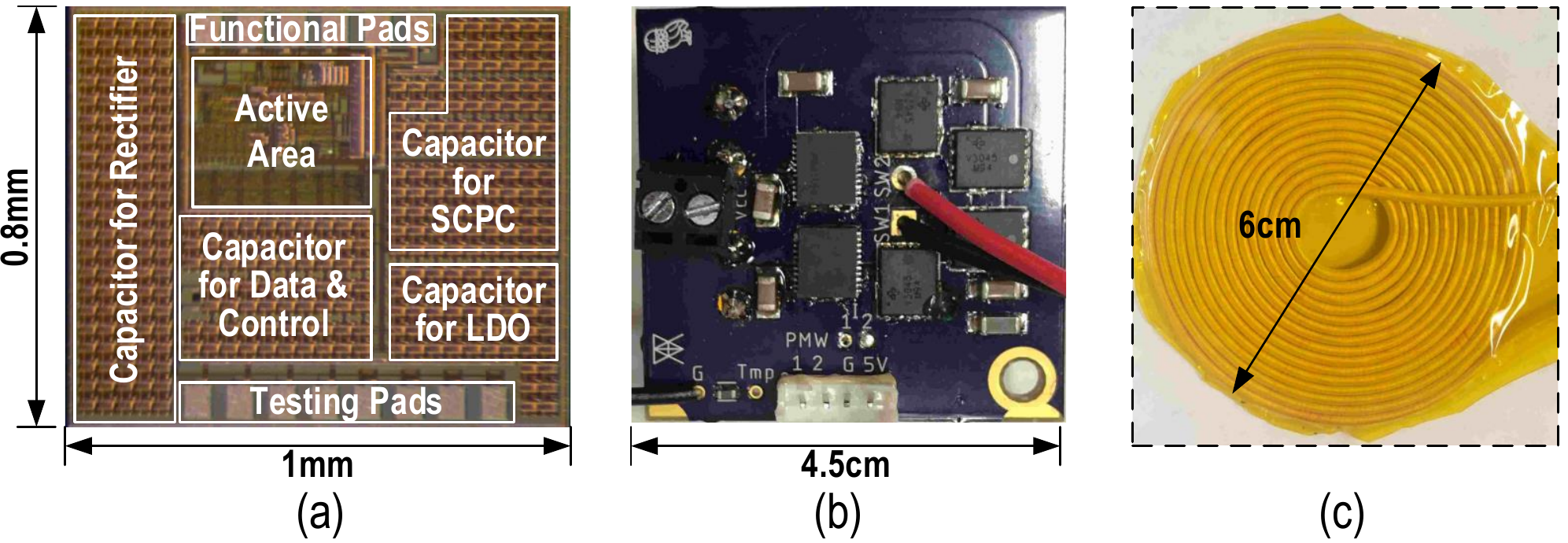}
      \caption{(a) Micrograph of the implant SoC, (b) the portable AC magnetic field driver and (c) the TX coil for the 330-kHz magnetic field generation.}
      \label{S4_Chip}
   \end{figure}
   
\begin{figure}[t]
      \centering
      \includegraphics[width=1\linewidth]{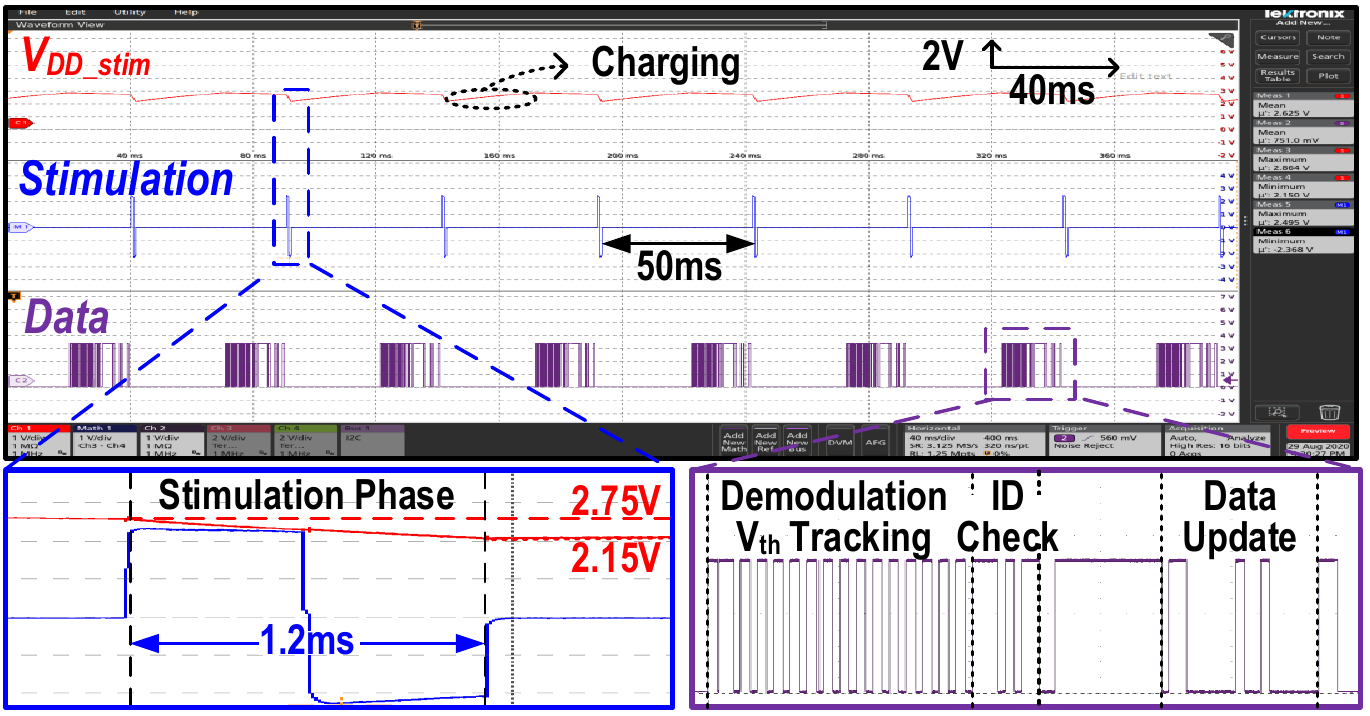}
      \caption{Measured operating waveform of the proposed implant.}
      \label{S4_Wave}
   \end{figure}

The on-chip 8-bit ID is realized with CMOS physical unclonable functions (PUF) instead of non-volatile memory. The PUF leverages transistor intrinsic variations to cheaply generate and store device-specific IDs. 
Fig.~\ref{S3_PUF}~(a) gives the design of the PUF cell adopting an inverter chain based topology and a native NMOS for the local supply regulation of 0.8~V \cite{li_251_2019}. 
In the simulation with 0.85-to-1.25-V $V_\mathrm{DD\_L}$, the PUF cell's virtual supply changes from 0.792 V to 0.807 V, demonstrating robustness against large supply voltage fluctuations.
Instead of using a level shifter, a skewed inverter built by a I/O PMOS and a core NMOS is utilized in the PUF cell to properly convert the voltage level within an ultra compact area.
Based on 5,000 simulation iterations, its output demonstrates same 50 percent of "0" and "1".
Thermal noise may flip the PUF cell's output and affect the bit stability.
As a solution to this, we adopt the temporal majority voting (TMV) technique in the 8-bit PUF circuit to ensure its reliability (Fig.~\ref{S3_PUF}~(b)). 
Because of the narrow operating temperature range required by the target applications and the adopted native voltage regulation, a 15-cycle TMV is sufficient to filter out the flipping error of each cell caused by thermal noise, as illustrated in Fig.~\ref{S3_PUF}~(c). 
Signal Y represents the final output of the PUF cell with the filtered flipping errors and will be sampled by $R_\mathrm{W}$ after each TMV cycle.
$R_\mathrm{W}$ also resets the counting circuitry that is shared by eight PUF cells to save the occupied area.
The ID generation is triggered by the POR signal at a steady supply, ensuring correct generation when the system turns on.
It is clock gated after the IDs are loaded to registers, making the power consumed by the PUF negligible in the entire system.

\section{Experimental Results}
\label{sec:measurement_result}

The implant’s SoC is fabricated in TSMC 180-nm CMOS technology with a 0.8~mm$^2$ area (Fig.~\ref{S4_Chip}~(a)).
We conduct all the tests with the custom TX consisting of a portable magnetic field driver (Fig.~\ref{S4_Chip}~(b)), a 6-cm 10-\si{\micro}H TX coil to generate the 330-kHz alternating magnetic field (Fig.~\ref{S4_Chip}~(c)), and a 12.3-cm$^3$ permanent magnet for the DC biasing magnetic field.

\begin{figure}[t]
      \centering
      \includegraphics[width=0.8\linewidth]{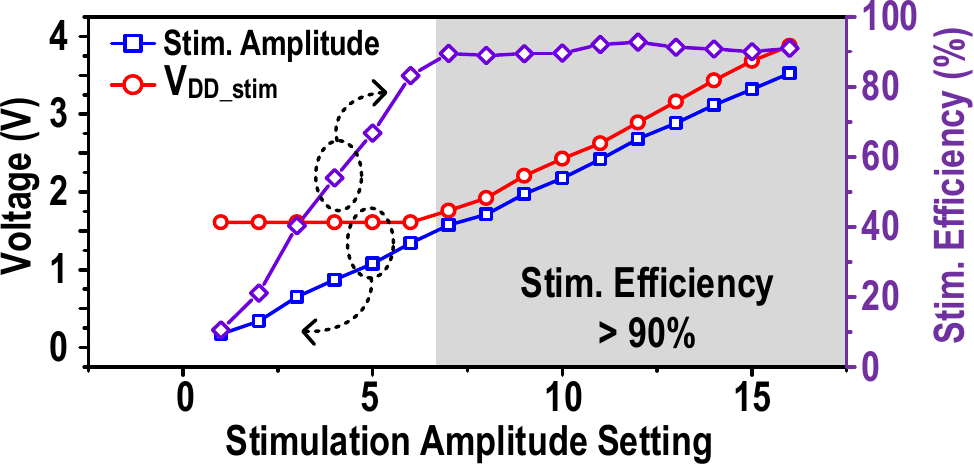}
      \caption{Measurement of programmable stimulation amplitudes and stimulator efficiency.}
      \label{S4_Stim}
   \end{figure}
   
\begin{figure}[t]
      \centering
      \includegraphics[width=0.85\linewidth]{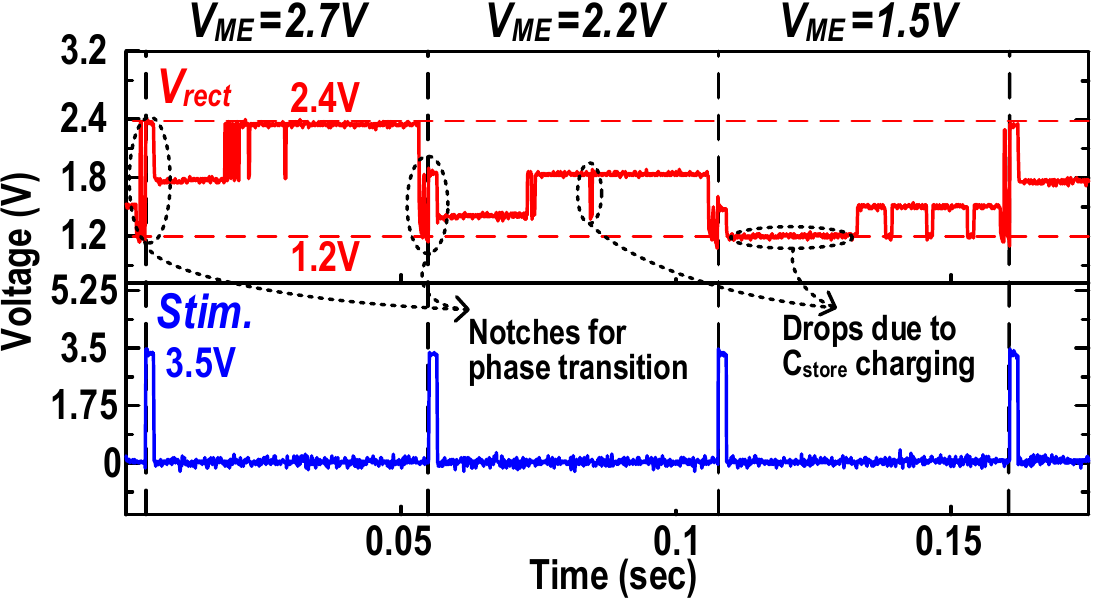}
      \caption{Operating robustness test with varying source voltage.}
      \label{S4_Vari}
   \end{figure}

\subsection{Functional Validations of the Proposed System}
\label{subsec:T_Ele}

Fig. \ref{S4_Wave} shows the measured operating waveform of the implant. The stimulation supply $V_\mathrm{DD\_stim}$ is charged up and regulated to 2.75~V by the adaptive voltage converter, then drops to 2.15~V after powering the 2.5-V, 1.2-ms biphasic stimulation pulse with 1-k\si{\Omega} load.
The implants demonstrate a wide programmable amplitude range from 0.25~V to 3.5~V with 4-bit resolution in Fig.~\ref{S4_Stim}. 
Thanks to the adaptive voltage conversion for the stimulation supply, 90\% stimulator efficiency (= (stimulation amplitude \si{\times} simulation current) / (stimulator supply \si{\times} simulator current)) is achieved when the stimulation amplitude is larger than 1.5~V.
The SoC merely consumes 9 \si{\micro}W without stimulation generation. 
When conducting high-power stimulation at therapeutic frequencies (20 - 200 Hz) \cite{singer_magnetoelectric_2020}, the entire system’s power is dominated by the average stimulation power (= $P_\mathrm{Stim}$ \si{\times} Pulse Width \si{\times} Frequency), resulting in maximum efficiency of around 90\% of the whole system.
To test the implant’s robustness against source variations, we intentionally change ME film voltage ($V_\mathrm{ME}$) by altering the magnetic field strength. It is verified by Fig.~\ref{S4_Vari} that the implant maintains its operation with maximum stimulation amplitude of 3.5~V under 1.5-to-3.5-V ME source voltage variations, demonstrating reliability of the system. 

\begin{figure}[t]
      \centering
      \includegraphics[width=0.95\linewidth]{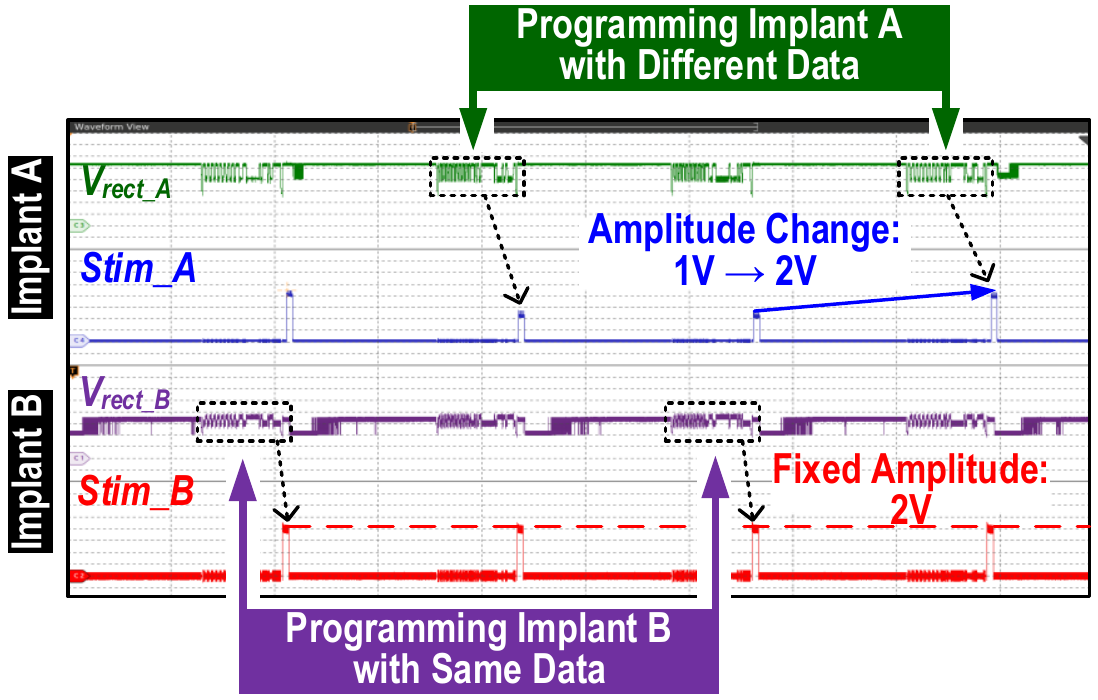}
      \caption{Multiple implants individually programmed by a single TX; the stimulation of implant A is updated while the stimulation of implant B is maintained.}
      \label{S4_Multi}
   \end{figure}
   
\begin{figure}[t]
      \centering
      \includegraphics[width=0.8\linewidth]{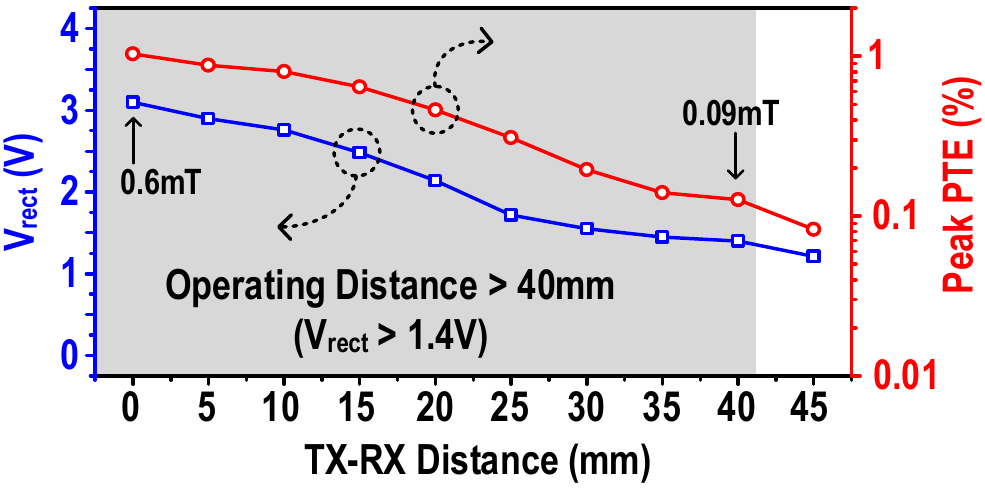}
      \caption{Test of implant at different distances from the TX. Measured received voltage and peak power transfer efficiency are reported.}
      \label{S4_Distance}
   \end{figure}
   
   \begin{figure}[t]
      \centering
      \includegraphics[width=1\linewidth]{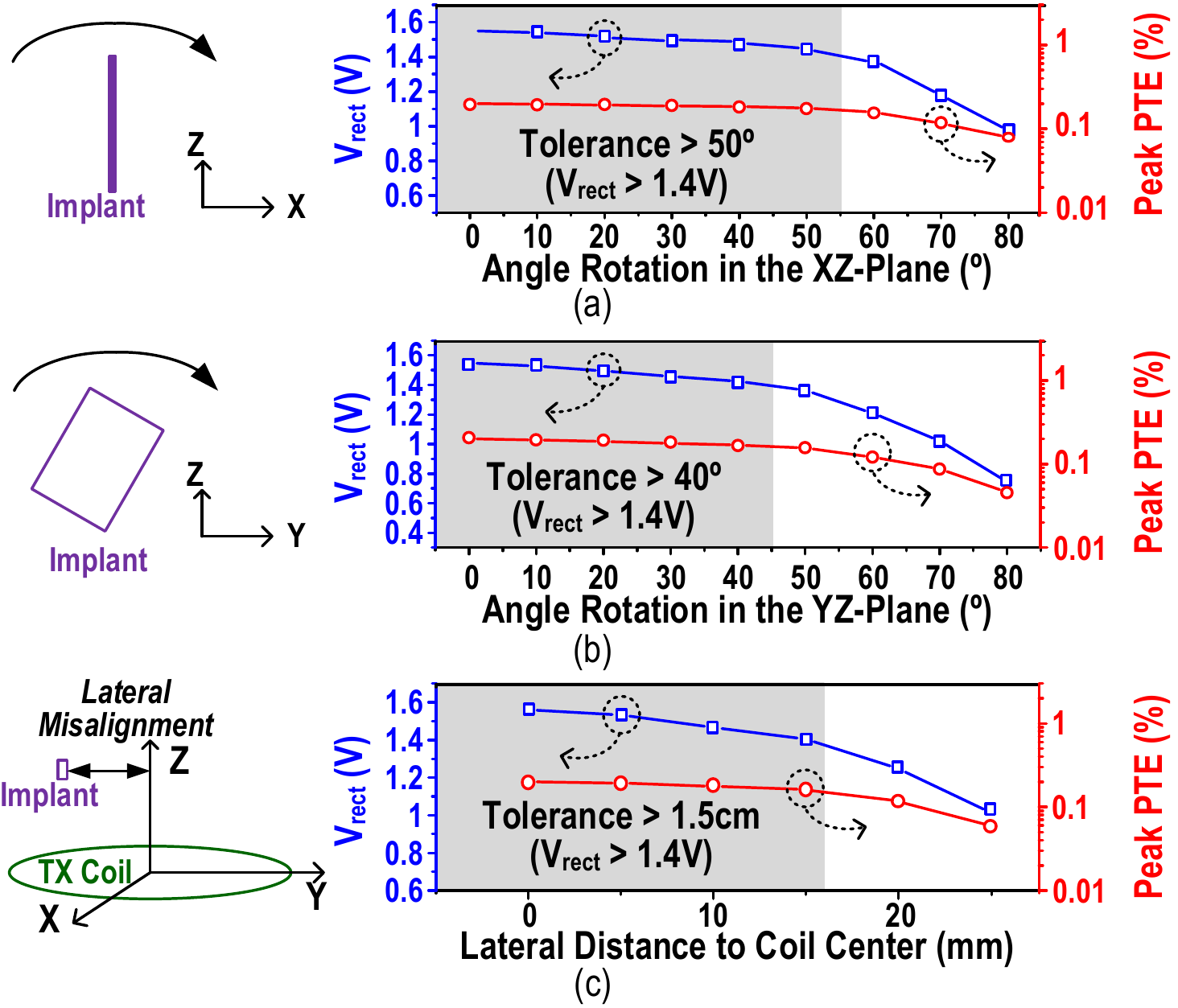}
      \caption{Tests of implant's operation with (a) angular misalignment in the XZ-plane, (b) angular misalignment in the YZ-plane and (c) lateral misalignment. All tests are conducted at 30 mm away from the TX.}
      \label{S4_Angel}
   \end{figure}

To validate the system's capabilities of individual addressing for coordinated stimulation, we use a single TX to power and program multiple implants placed at different distances of 15 mm and 25 mm. Stimulating patterns of the implants A and B are separately programmed in different operating cycles by the downlink data containing corresponding device IDs (11010111 for implant A and 01111000 for implant B). Fig.~\ref{S4_Multi} shows that stimulation amplitude of the implant A is changed from 1~V to 2~V without affecting the stimulation generated by the implant B. Since data transferred in the allocated cycles for the implant B does not alter, stimulation of the implant B maintains a voltage of 2~V as expected.
In this procedure, powering multiple implants does not add extra power consumption to the TX, which means higher system efficiency can be reached by involving more devices.  

To account for the uncertainties in implantation, such as variances in device placement during surgery and tissue movement over time, the implants are supposed to be robust against varying TX-RX distance and misalignment.
In order to assess this, we test the devices at different distances from the TX with angular and lateral misalignment, as shown in Fig.~\ref{S4_Distance} and Fig.~\ref{S4_Angel}. 
Since perturbations in distance and alignments can affect the power link stability, the received voltage $V_\mathrm{rect}$ is reported here as an indicator, whose insufficiency may cause failures in operation. Plots of measured PTE in these cases are included as well for channel quality evaluation.
As shown by Fig.~ \ref{S4_Distance}, the implants can reliably operate with a source amplitude as low as 1.5 V, demonstrating tolerance of 40-mm TX-implant distance change; The highest power transfer efficiency of 1.03\% is achieved when the implant is at the center of the TX coil with ideal alignment. 
In tests of angular misalignment sensitivity that are conducted at 30 mm from the TX coil, contributed by ME and robust SoC design, the proposed devices demonstrate robustness against a 50-degree angle rotation in the XZ-plane (Fig.~\ref{S4_Angel}~(a)) and a 40-degree angle rotation in the YZ-plane (Fig.~\ref{S4_Angel}~(b)).
Additionally, Fig.~\ref{S4_Angel}~(c) demonstrates that the implantable devices are capable of operating with a 1.5-cm lateral misalignment, when the TX-implant distance is 3~cm.

\begin{figure}[t]
      \centering
      \includegraphics[width=0.85\linewidth]{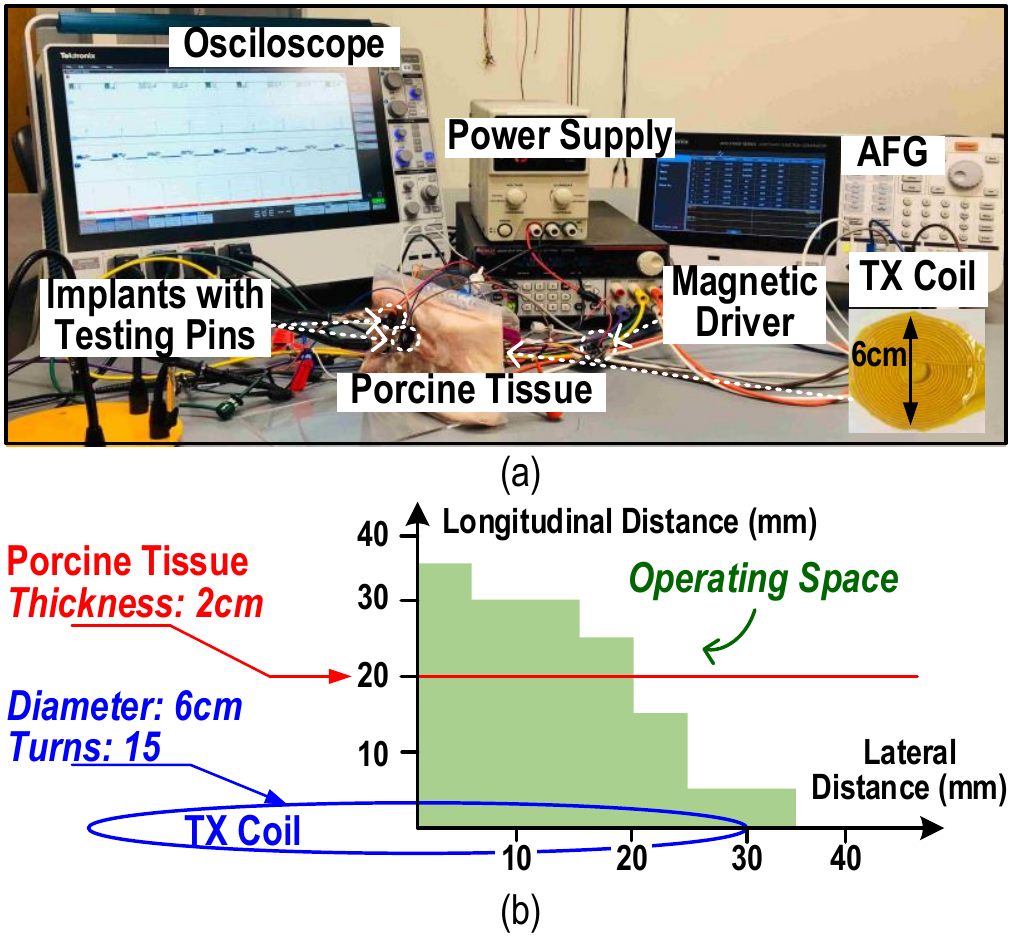}
      \caption{(a) \textit{In-vitro} test setup and (b) shmoo plot showing the operating space with the porcine tissue.}
      \label{S4_Porcine}
   \end{figure}

\subsection{In-Vitro Tests and Safety Analysis}
\label{subsec:T_Vitro} 

\begin{figure}[t]
      \centering
      \includegraphics[width=0.9\linewidth]{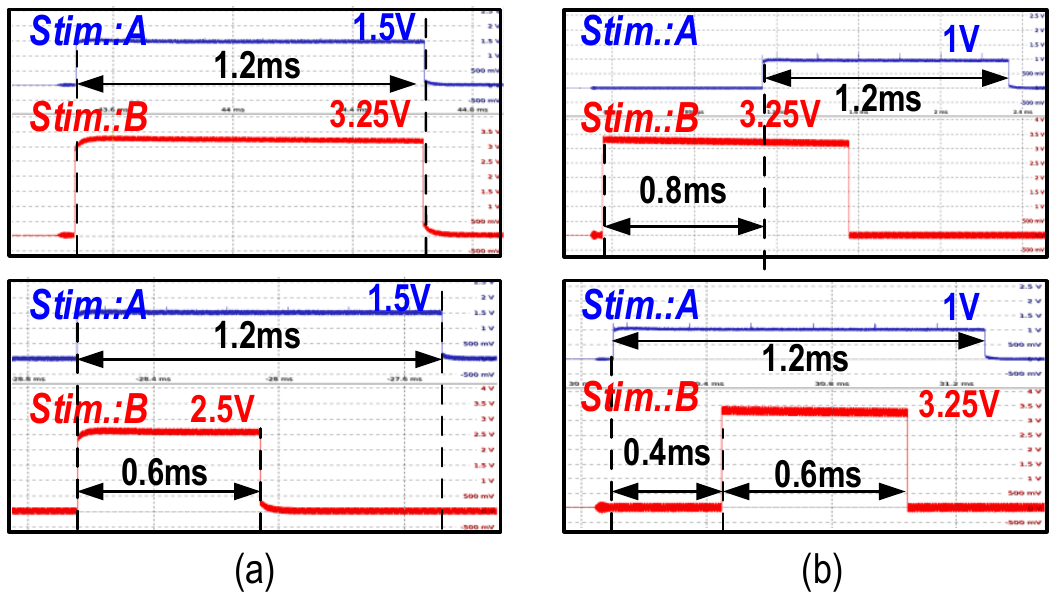}
      \caption{Measured (a) synchronized stimulation without delay and (b) stimulation with programmable delays of two implants \textit{in-vitro}.}
      \label{S4_Vitro}
   \end{figure}

To evaluate the proposed devices' performance in biological tissue, we performed \textit{in-vitro} tests with a 2-cm thick porcine tissue as a medium (Fig.~\ref{S4_Porcine}~(a)). 
The implants can receive sufficient power transmitted through the porcine tissue and reliably operate with a TX-implant separation up to 3.5 cm (there is a gap of air between the TX and the implants when the longitudinal distance is greater than 2 cm). 
In addition, they can operate at a 3.5-cm lateral distance if the ME films are close to the TX coil (around 5mm) and tolerate a 1.5-cm lateral misalignment when they are 3 cm away from the TX (Fig.~\ref{S4_Porcine}~(b)).
Synchronized stimuli of multiple implants powered by a shared TX with programmable amplitudes, pulse width, and start delays are also demonstrated \textit{in-vitro}, as given by Fig.~\ref{S4_Vitro}.
   
\begin{figure}[t]
      \centering
      \includegraphics[width=1\linewidth]{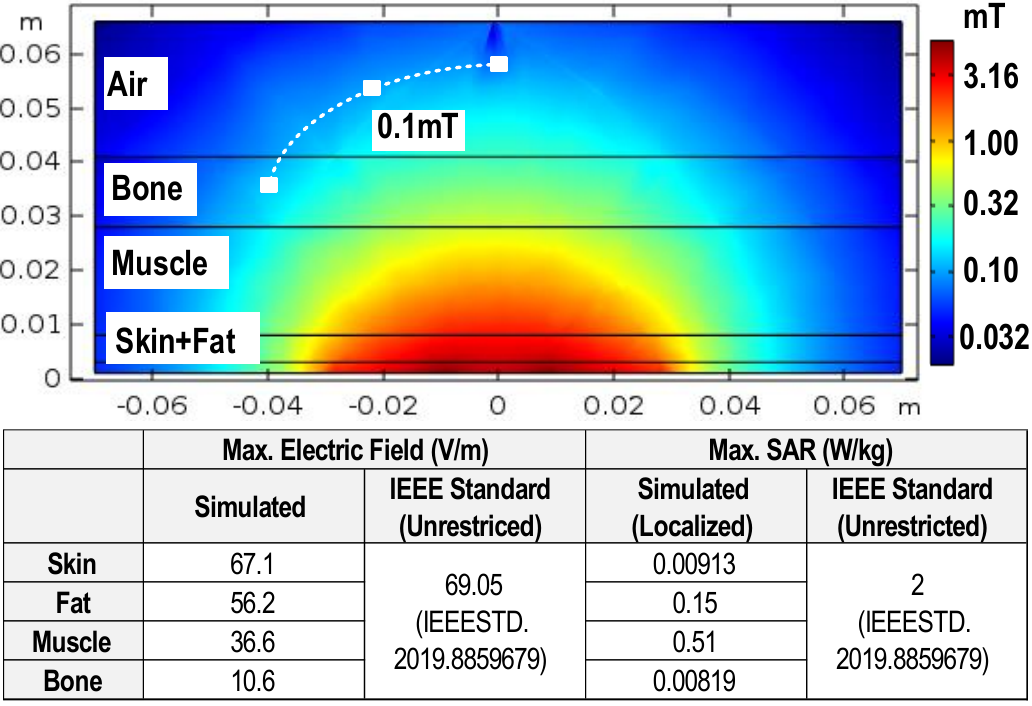}
      \caption{Local safety analysis of the coil-generated 330-kHz magnetic field in COMSOL, a 0.1-mT field strength at 60-mm depth is achieved with safety constrains.}
      \label{S4_Safe}
   \end{figure}
   
\begin{figure}[t]
      \centering
      \includegraphics[width=0.95\linewidth]{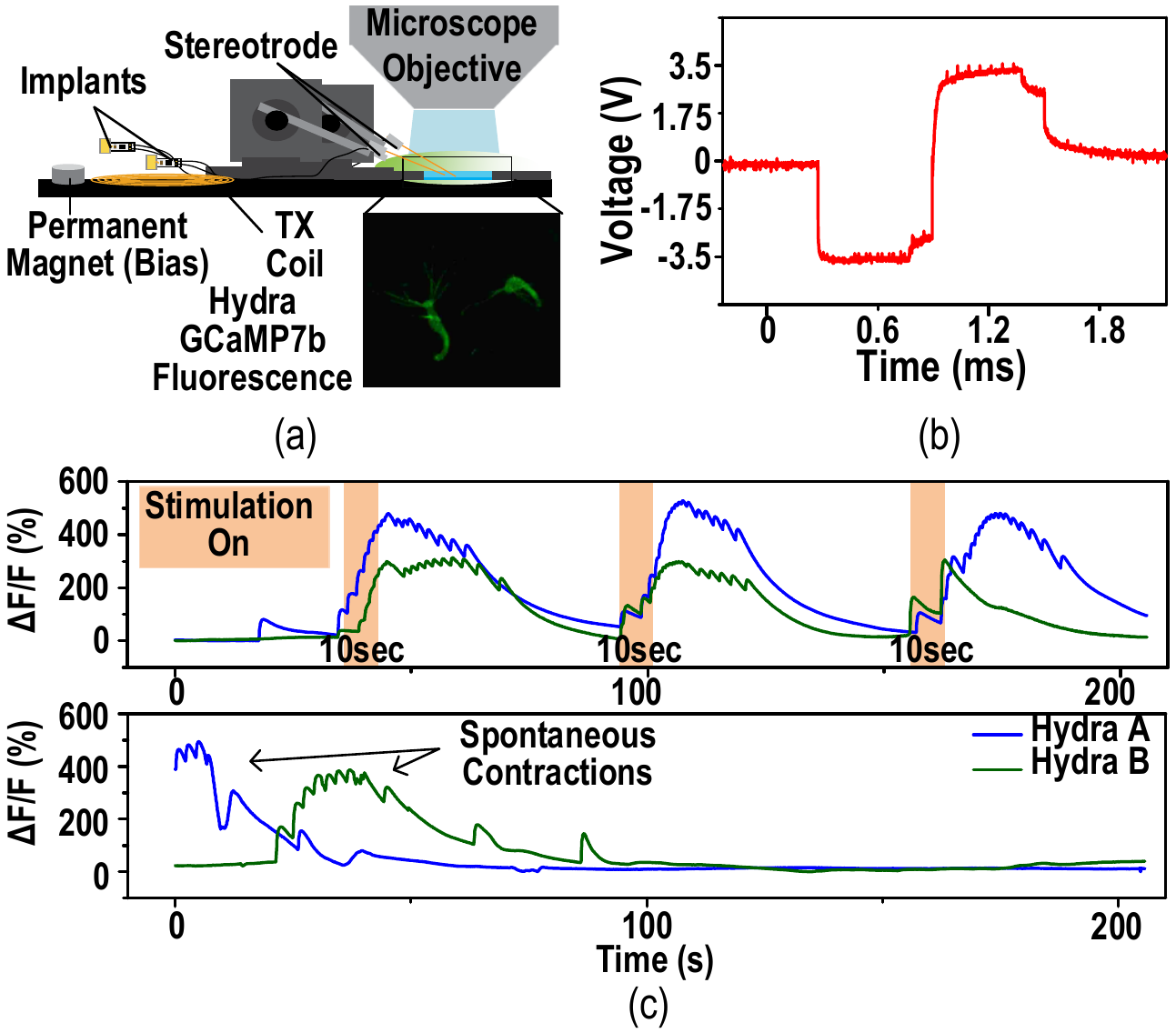}
      \caption{(a) Experimental setup of the Hydra test, (b) waveform of the applied stimulation pulse and (c) synchronous muscle activation in response to the electrical stimuli.}
      \label{S4_Hydra}
   \end{figure}

To analyze the EM exposure safety of the system for implantation, the specific absorption rate and the electric field induction in a coil-generated 330~kHz magnetic field are studied in COMSOL.
A multiple-layer human tissue model consisting of skin, fat, muscle and bone is built for this analysis. 
With the coil dimension in our TX, a magnetic strength of 0.1~mT, which is enough to sustain implant’s functionality, can be achieved at 60~mm without violating IEEE safety standards in the body model (Fig~\ref{S4_Safe}). In our experiments, we use a smaller power TX and demonstrate a maximum TX-implant distance of 40~mm in air with margins on safety limits. This simulation result also suggests that larger distance up to 60~mm is feasible, which could be useful for applications requiring large implantation depth, such as the cardiac pacing.

\begin{table*}[t]
\caption{\textbf{Comparisons with State-of-the-Art Biomedical Stimulating System and Commercial Stimulators}}
\label{table}
\centering
\setlength{\tabcolsep}{3.6pt}
\renewcommand{\arraystretch}{1.5}
\begin{tabular}{|p{70pt}|p{55pt}|p{40pt}|p{40pt}|p{45pt}|p{45pt}|p{45pt}|p{50pt}|p{50pt}|}
\hline

& 
\parbox[c][1.2cm]{55pt}{\centering
{\textbf{This \\ Work}}}&
\parbox[c][1.2cm]{40pt} {\centering{ISSCC 16 \\ \cite{lo_176-channel_2016}}}&
\parbox[c][1.2cm]{40pt} {\centering{ISSCC 18 \\ \cite{jia_mm-sized_2018}}}&
\parbox[c][1.2cm]{45pt}{\centering
{Nat. \\ Commun. 19 \\ \cite{gutruf_wireless_2019}}}&
\parbox[c][1.2cm]{45pt} {\centering{Scientific \\ Reports 20 \\ \cite{lyu_synchronized_2020}}} &
\parbox[c][1.2cm]{45pt} {\centering{Nat. Biomed.\\Eng. 20 \\ \cite{piech_wireless_2020}}} &
\parbox[c][1.2cm]{50pt} {\centering{Commercial \\ Leadless \\ Pacemaker~\cite{medtronic_micra_nodate}}} 
& 
\parbox[c][1.2cm]{50pt}{\centering{Commercial \\ Dual-Chamber \\Pacemaker~\cite{noauthor_accolade_nodate}}} 
\\\hline

\parbox[c][0.8cm]{70pt}{\centering
{\textbf{Target \\ Application}}} &
\parbox[c][0.8cm]{55pt}{\centering
{\textbf{Spinal Cord, \\ Cardiac}}}&
\parbox[c][0.8cm]{40pt} {\centering{Spinal \\ Cord}}&
\parbox[c][0.8cm]{40pt} {\centering{Brain}}&
\parbox[c][0.8cm]{45pt}{\centering
{Cardiac}}&
\parbox[c][0.8cm]{45pt} {\centering{Cardiac}} &
\parbox[c][0.8cm]{45pt} {\centering{Peripheral Nerve}} &
\parbox[c][0.8cm]{50pt} {\centering{Cardiac}} 
&
\parbox[c][0.8cm]{50pt} {\centering{Cardiac}} 
\\\hline

\multicolumn{1}{|c|}{\textbf{Process (nm)}} & 
\multicolumn{1}{c|}{\textbf{180}}& 
\multicolumn{1}{c|}{180}&
\multicolumn{1}{c|}{350}&
\multicolumn{1}{c|}{N/A}& 
\multicolumn{1}{c|}{180}&
\multicolumn{1}{c|}{65}&
\multicolumn{1}{c|}{N/A}&
\multicolumn{1}{c|}{N/A}
\\\hline

\multicolumn{1}{|c|}{\textbf{Power Source}} 
& \multicolumn{1}{c|}{\textbf{Magnetoelectric}}
& \multicolumn{1}{c|}{Inductive}
& \multicolumn{1}{c|}{Inductive} 
& \multicolumn{1}{c|}{Inductive}
& \multicolumn{1}{c|}{Inductive} 
& \multicolumn{1}{c|}{Ultrasonic}
& \multicolumn{1}{c|}{Battery}
& \multicolumn{1}{c|}{Battery}
\\\hline

\multicolumn{1}{|c|}{\textbf{Carrier Freq. (MHz)}} 
& \multicolumn{1}{c|}{\textbf{0.33}}
& \multicolumn{1}{c|}{2}
& \multicolumn{1}{c|}{60}
& \multicolumn{1}{c|}{13.56}
& \multicolumn{1}{c|}{13.56, 40.68}
& \multicolumn{1}{c|}{1.85}
& \multicolumn{1}{c|}{N/A}
& \multicolumn{1}{c|}{N/A}
\\\hline


\parbox[c][0.8cm]{70pt}{\centering
{\textbf{Multisite \\ Stim. Strategy}}} &
\parbox[c][0.8cm]{55pt}{\centering
{\textbf{Single TX + \\ Multi. Implants}}}&
\parbox[c][0.8cm]{40pt} {\centering{Electrode \\ Array}}&
\parbox[c][0.8cm]{40pt} {\centering{\si{\micro}LED \\ Array}}&
\parbox[c][0.8cm]{45pt}{\centering
{Wired \\ Electrodes}}&
\parbox[c][0.8cm]{45pt} {\centering{Two TXs + \\ Two Implants}} &
\parbox[c][0.8cm]{45pt} {\centering{N/A}} &
\parbox[c][0.8cm]{50pt} {\centering{N/A}} &
\parbox[c][0.8cm]{50pt} {\centering{Wired \\ Electrodes}} 
\\\hline

\parbox[c][0.8cm]{70pt}{\centering
{\textbf{Individual \\ Addressability}}} &
\parbox[c][0.8cm]{55pt}{\centering
{\textbf{PUF ID}}}&
\parbox[c][0.8cm]{40pt} {\centering{Multiplexer}}&
\parbox[c][0.8cm]{40pt} {\centering{Multiplexer}}&
\parbox[c][0.8cm]{45pt}{\centering
{Multiplexer}}&
\parbox[c][0.8cm]{45pt} {\centering{Different \\ Carrier Freq.}} &
\parbox[c][0.8cm]{45pt} {\centering{N/A}} &
\parbox[c][0.8cm]{50pt} {\centering{N/A}} &
\parbox[c][0.8cm]{50pt} {\centering{N/A}} 
\\\hline

\multicolumn{1}{|c|}{\textbf{Stimulation Type}} 
& \multicolumn{1}{c|}{\textbf{Voltage}}
& \multicolumn{1}{c|}{Current}
& \multicolumn{1}{c|}{Optical} 
& \multicolumn{1}{c|}{Voltage}
& \multicolumn{1}{c|}{Voltage} 
& \multicolumn{1}{c|}{Current}
& \multicolumn{1}{c|}{Voltage}
& \multicolumn{1}{c|}{Voltage}
\\\hline

\parbox[c][0.8cm]{70pt}{\centering
{\textbf{Max. Stim. Amp.\\ / Resolution}}} &
\parbox[c][0.8cm]{55pt}{\centering
{\textbf{3.5 V / \\ 4 bits}}}&
\parbox[c][0.8cm]{40pt} {\centering{0.5 mA / \\ 7 bits}}&
\parbox[c][0.8cm]{40pt} {\centering{N/A}}&
\parbox[c][0.8cm]{45pt}{\centering
{3 V / \\ Fixed}}&
\parbox[c][0.8cm]{45pt} {\centering{3 V / \\ Fixed}} &
\parbox[c][0.8cm]{45pt} {\centering{0.4 mA / \\ 3 bits}} &
\parbox[c][0.8cm]{50pt} {\centering{5 V / \\ 40 Levels}} &
\parbox[c][0.8cm]{50pt} {\centering{7.5 V / \\ N/A}} 
\\\hline

\parbox[c][0.8cm]{70pt}{\centering
{\textbf{Max. Pulse Width \\ / Resolution}}} &
\parbox[c][0.8cm]{55pt}{\centering
{\textbf{1.2 ms / \\ 4 bits}}}&
\parbox[c][0.8cm]{40pt} {\centering{8 ms / \\ Fixed}}&
\parbox[c][0.8cm]{40pt} {\centering{2 ms / \\ 2 bits}}&
\parbox[c][0.8cm]{45pt}{\centering
{5 ms / \\ Fixed}}&
\parbox[c][0.8cm]{45pt} {\centering{0.3 ms / \\ 3 Levels}} &
\parbox[c][0.8cm]{45pt} {\centering{Continuous}} &
\parbox[c][0.8cm]{50pt} {\centering{1 ms / \\ 5 Levels}} &
\parbox[c][0.8cm]{50pt} {\centering{2 ms / \\ N/A}} 
\\\hline

\parbox[c][0.8cm]{70pt}{\centering
{\textbf{Max. Delay \\ / Resolution}}} &
\parbox[c][0.8cm]{55pt}{\centering
{\textbf{0.8 ms / \\ 5 bits}}}&
\parbox[c][0.8cm]{40pt} {\centering{N/A}}&
\parbox[c][0.8cm]{40pt} {\centering{N/A}}&
\parbox[c][0.8cm]{45pt}{\centering
{N/A}}&
\parbox[c][0.8cm]{45pt} {\centering{N/A}} &
\parbox[c][0.8cm]{45pt} {\centering{N/A}} &
\parbox[c][0.8cm]{50pt} {\centering{N/A}} &
\parbox[c][0.8cm]{50pt} {\centering{N/A}} 
\\\hline

\multicolumn{1}{|c|}{\textbf{SoC Power (\si{\micro}W)}}  
& \multicolumn{1}{c|}{\textbf{9}} 
& \multicolumn{1}{c|}{864} 
& \multicolumn{1}{c|}{300}
& \multicolumn{1}{c|}{N/A}
& \multicolumn{1}{c|}{3}
& \multicolumn{1}{c|}{4}
& \multicolumn{1}{c|}{N/A}
& \multicolumn{1}{c|}{N/A}
\\\hline

\multicolumn{1}{|c|}{\textbf{Chip Area (mm$^2$)}}  
& \multicolumn{1}{c|}{\textbf{1 x 0.8}} 
& \multicolumn{1}{c|}{5.7 x 4.4} 
& \multicolumn{1}{c|}{1 x 1}
& \multicolumn{1}{c|}{N/A}
& \multicolumn{1}{c|}{0.85 x 0.45}
& \multicolumn{1}{c|}{1 x 1}
& \multicolumn{1}{c|}{N/A}
& \multicolumn{1}{c|}{N/A}
\\\hline

\multicolumn{1}{|c|}{\textbf{Implant Size (mm$^3$)}}  
& \multicolumn{1}{c|}{\textbf{6.2}} 
& \multicolumn{1}{c|}{500} 
& \multicolumn{1}{c|}{12.2}
& \multicolumn{1}{c|}{N/A}
& \multicolumn{1}{c|}{10.1}
& \multicolumn{1}{c|}{1.7}
& \multicolumn{1}{c|}{800}
& \multicolumn{1}{c|}{16050}
\\\hline

\parbox[c][0.8cm]{70pt}{\centering
{\textbf{Max. TX-Implant \\ Distance (mm)}}} &
\parbox[c][0.8cm]{55pt}{\centering
{\textbf{40}}}&
\parbox[c][0.8cm]{40pt} {\centering{N/A}}&
\parbox[c][0.8cm]{40pt} {\centering{7}}&
\parbox[c][0.8cm]{45pt}{\centering
{30}}&
\parbox[c][0.8cm]{45pt} {\centering{60}} &
\parbox[c][0.8cm]{45pt} {\centering{66 (Min. \\ Distance: 42)}} &
\parbox[c][0.8cm]{50pt} {\centering{N/A}} &
\parbox[c][0.8cm]{50pt} {\centering{N/A}} 
\\\hline

\parbox[c][0.8cm]{70pt}{\centering
{\textbf{Max. Distance \\ / Implant Size}}} &
\parbox[c][0.8cm]{55pt}{\centering
{\textbf{6.45}}}&
\parbox[c][0.8cm]{40pt} {\centering{N/A}}&
\parbox[c][0.8cm]{40pt} {\centering{0.57}}&
\parbox[c][0.8cm]{45pt}{\centering
{N/A}}&
\parbox[c][0.8cm]{45pt} {\centering{5.94}} &
\parbox[c][0.8cm]{45pt} {\centering{38.82}} &
\parbox[c][0.8cm]{50pt} {\centering{N/A}} &
\parbox[c][0.8cm]{50pt} {\centering{N/A}} 
\\\hline


\parbox[c][1.1cm]{70pt}{\centering
{\textbf{Angular \& Lateral \\ Misalignment \\ Tolerance}}} &
\parbox[c][1.1cm]{55pt}{\centering
{\textbf{50$^{\circ}$, 15 mm 
}}}&
\parbox[c][1.1cm]{40pt} {\centering{N/A}}&
\parbox[c][1.1cm]{40pt} {\centering{N/A}}&
\parbox[c][1.1cm]{45pt}{\centering
{N/A}}&
\parbox[c][1.1cm]{45pt} {\centering{N/A}} &
\parbox[c][1.1cm]{45pt} {\centering{45$^{\circ}$, 2.2 mm 
}} &
\parbox[c][1.1cm]{50pt} {\centering{N/A}} &
\parbox[c][1.1cm]{50pt} {\centering{N/A}} 
\\\hline

\end{tabular}
\label{tab2}
\end{table*}

\begin{figure}[t]
      \centering
      \includegraphics[width=0.98\linewidth]{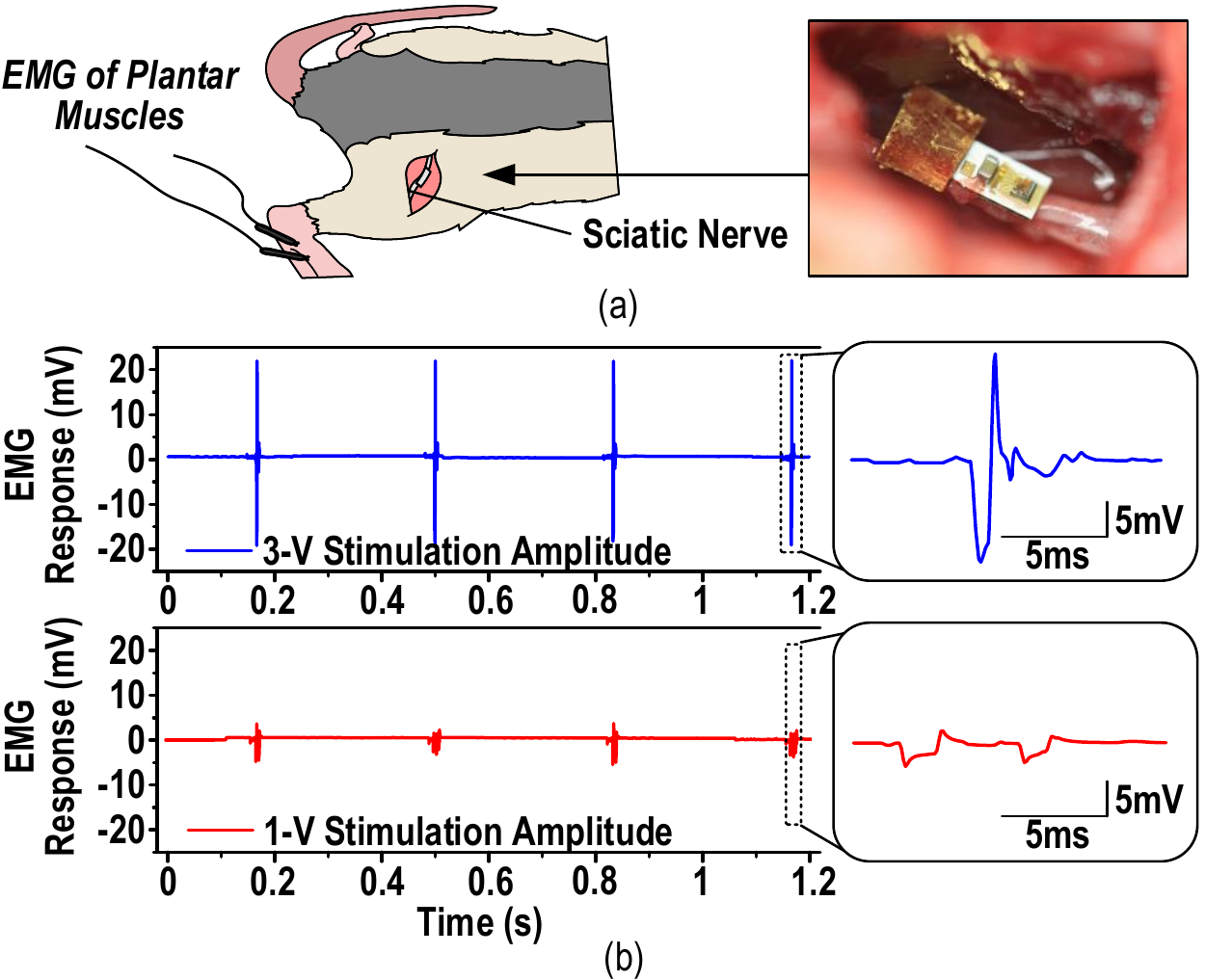}
      \caption{(a) Illustration of the rat experiment and (b) EMG responses of the rat with 3-V and 1-V stimulation amplitude.}
      \label{S4_Rat}
   \end{figure}
   
\subsection{In-Vivo Experiments}
\label{subsec:T_Vivo}  

\subsubsection{Hydra Experiments for Muscle Stimulation}
To further assess the system’s bio-stimulation capability, we test this work with \textit{Hydra vulgaris} as a model for coordinated muscle stimulation, using the experimental setup shown in Fig.~\ref{S4_Hydra}~(a).
Considering the millimeter size of these animals, we use stereotrodes attached to the device to precisely deliver stimuli to individual animals. While \textit{Hydra} naturally express voltage-gated ion channels making their tissue electrically excitable \cite{siebert_stem_2019}, we use a transgeneic line that expresses a calcium-sensitive fluorescent protein (GCaMP7b) to visualize stimulation of epithelial muscle cells \cite{badhiwala_multiple_2021}. 
To model synchronous stimulation of muscle tissue as is necessary in biventricular cardiac pacing, we use two nearby animals to demonstrate simultaneous stimulation of motes at varying heights with a single transmitting coil. 
\textit{Hydra} are known to have spontaneous and often asynchronous contractions. In order to synchronize the muscle contractions, herein, we provide 3.5~V, 20~Hz, 1.2~ms pulse width, biphasic stimulation pulse trains to synchronize the \textit{Hydra}'s muscle contractions (Fig.~\ref{S4_Hydra}~(b)). The synchronous stimuli results in GCaMP7b fluorescence increases greater than 200\% that demonstrates activation of the voltage-gated ion channels resulting in stimulus aligned epitheliomuscle contractions at the same time in both organisms, as shown in Fig.~\ref{S4_Hydra}~(c).

\subsubsection{Rat Experiments for Nerve Stimulation}
The device is also validated \textit{in vivo} with rat as a model for nerve stimulation (Fig.~\ref{S4_Rat}~(a)). 
The rat is anesthetized using 5\% isoflurane in oxygen before being transferred to a heated pad and 2\% isoflurane in oxygen for the subsequent surgery to expose the sciatic nerve.
Two electromyography (EMG) needle electrodes are placed in the rat's foot with a third reference needle placed in the shoulder region to record any resulting muscle activity.
The TX is placed out of the body and aligned to the implanted device at around 2 cm.
The implant is encapsulated with parylene C coating.
To demonstrate programmability, varying charge injection can allow stimulators to recruit a different number of muscle motor units. 
We apply 1.2 ms, 3 Hz biphasic stimulation with different amplitudes through the wirelessly powered implanted device to the rat's sciatic nerve and then observe and record any resulting leg kicks and corresponding EMG signals.
When various stimulation voltages are applied, we observe graded responses in the intensity of the rat leg kick with EMG of the plantar muscles. 
As shown in Fig.~\ref{S4_Rat}~(b), the EMG response with 3-V stimulation pulses shows a much stronger magnitude than that with 1-V stimulation pulses.

\subsection{Performance Summary and Comparison}
\label{subsec:Comp}

As given in Table~\ref{tab2}, in comparison with state-of-the-art for biomedical stimulation  \cite{lo_176-channel_2016, jia_mm-sized_2018, gutruf_wireless_2019, lyu_synchronized_2020, piech_wireless_2020}, this work proposes a novel multisite stimulation strategy of using a single TX to power and control multiple implants, which simultaneously achieves flexible deployment of stimuli, leadless structure, synchronized device operation and highly scalable channel quantity. 
Each implant is individually programmed through ID check with the PUF IDs and efficiently generates stimulation with fully programmable parameters covering amplitude, frequency, pulse width and delay.
In comparing the figure-of-merit of the ratio between the maximum distance and the implant volume \cite{piech_wireless_2020}, the proposed ME implants show a better value than these inductively powered state-of-the-arts \cite{jia_mm-sized_2018, lyu_synchronized_2020}.
While the ultrasonic device can operate at a larger distance, it requires a minimum TX-implant separation of around 4.2 cm \cite{piech_wireless_2020}, tolerating smaller distance variations and probably limiting its applications.
In addition, \cite{piech_wireless_2020} shows a higher sensitivity to lateral misalignment than the ME device.
Compared to the mostly advanced commercial pacemakers \cite{medtronic_micra_nodate, noauthor_accolade_nodate}, the proposed devices show comparable pacing capability with batteryless structure and significantly smaller volume and weight, making them more advantaged in wireless coordinated cardiac pacing. 

\section{Conclusion}
In conclusion, this work presents a proof-of-concept demonstration of multiple millimeter-sized implants remotely powered and controlled by a single TX for coordinated multisite bio-stimulation. 
Compared to previous work exploiting integrated electrode array, wired electrodes or frequency multiplexed TXs-implants pairs, the proposed novel single-TX multiple-implant strategy realizes more flexible stimuli deployment, easier synchronization, higher efficiency and improved scalability.
Magnetoelectric wireless power and data link is adopted and optimized in this work, because of its good efficiency under size constrains, low misalignment sensitivity and low tissue absorption in power delivery.
The robust and efficient SoC design enables the implants to operate reliably with a 2-V source amplitude change, perform individual programming through PUF IDs, and generate synchronized stimulation with a maximum efficiency of 90\% and fully programmable patterns. 
The implantable devices hold robustness against 50-degree angular misalignment and 1.5-cm lateral misalignment at a 30-mm implantation depth. 
These key features bring great advantages to the proposed work for clinical applications.

\appendices


\section*{Acknowledgment}

We would like to thank C. S. Edwin Lai for his surgical assistance and teaching of the rat peripheral nerve surgery. We also appreciate the culturing and handling of Hydra and useful discussions of Hydra biology from Soonyoung Kim.

\ifCLASSOPTIONcaptionsoff
  \newpage
\fi




\bibliographystyle{ieeetr}
%



%

\begin{IEEEbiography}[{\includegraphics[width=1in,height=1.25in,clip,keepaspectratio]{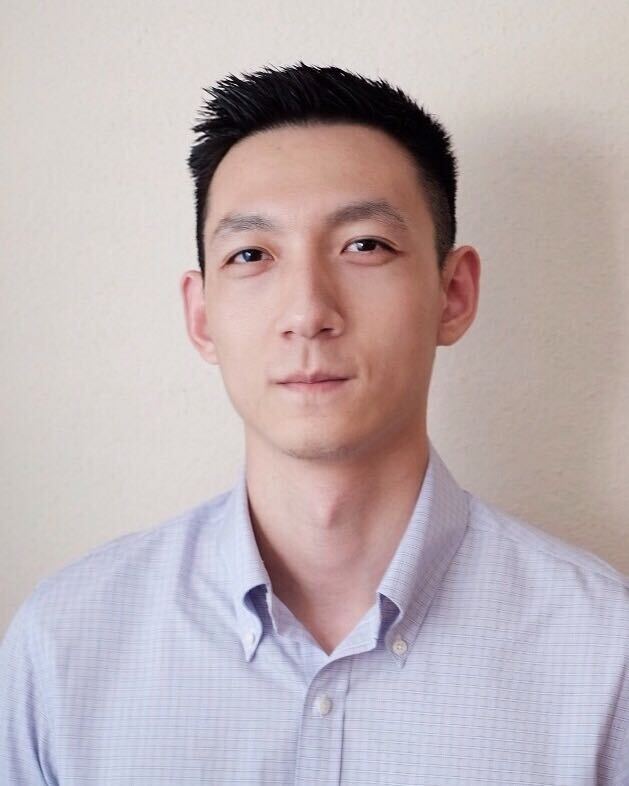}}]{Zhanghao Yu} received the B.E. degree in Integrated Circuit Design and Integrated System from the University of Electronic Science and Technology of China, Chengdu, China, in 2016, and the M.S. degree in Electrical Engineering from the University of Southern California, Los Angeles, CA, in 2018. He is currently working toward his Ph.D. degree in Electrical and Computer Engineering at Rice University, Houston, TX. He received the Best Paper Award at the 2021 IEEE Custom Integrated Circuits Conference (CICC).

His research interests include analog and mixed-signal integrated circuits design for bio-electronics, wireless power transfer and power management, low-power communication, and hardware security.
\end{IEEEbiography}

\begin{IEEEbiography}[{\includegraphics[width=1in,height=1.25in,clip,keepaspectratio]{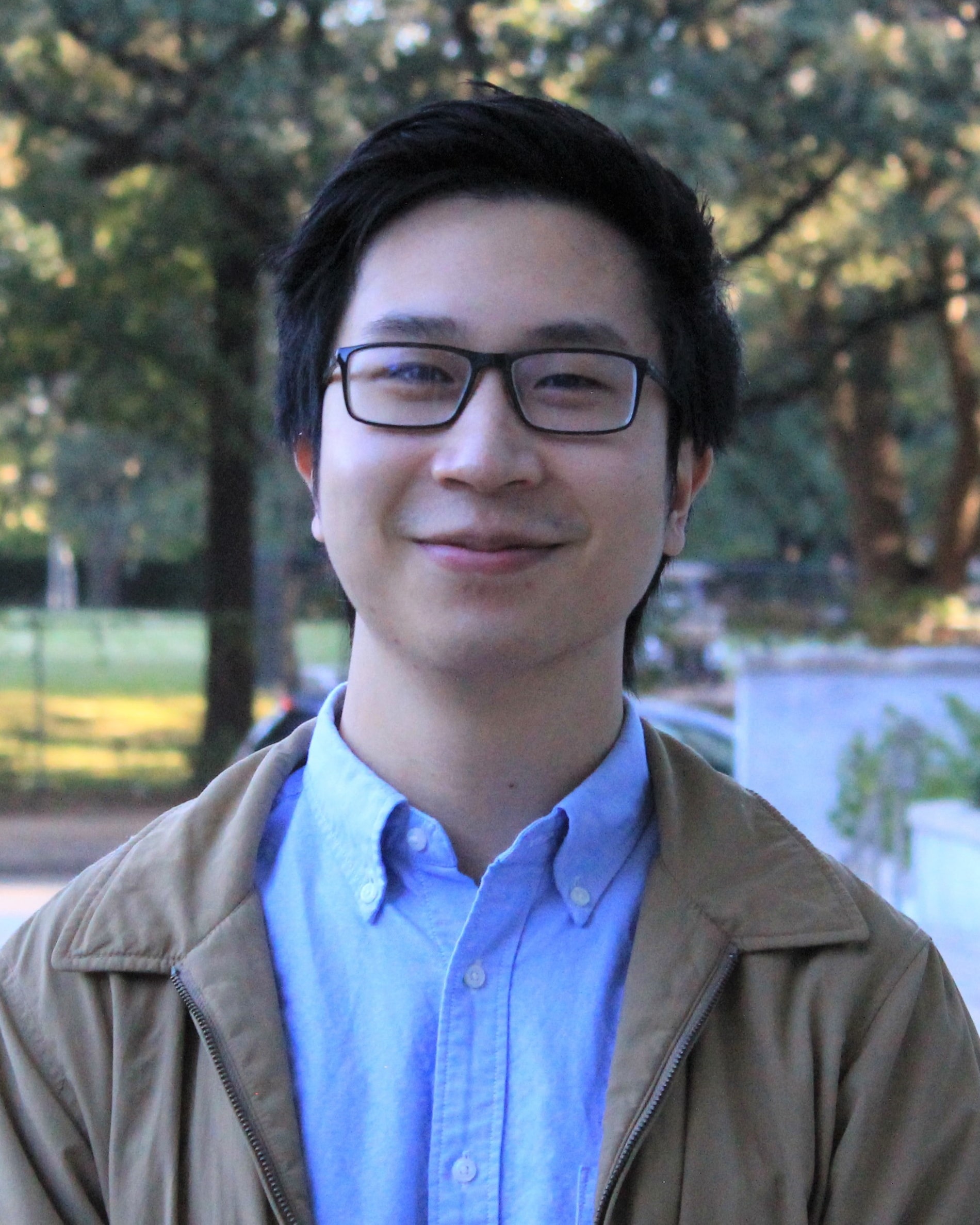}}]{Joshua Chen} received the B.S. degree in Bioengineering from the University of California, Berkeley in 2016. During his time at Berkeley, he was a member of the Berkeley Sensors and Actuators Center (BSAC) where he worked on developing technology for biomedical sensors and 3D printed microfluidics. After graduation, he worked at Verily Life Sciences prior to attending Rice University in Houston, TX to pursue the Ph.D. degree in Bioengineering.

His research interests are in neural engineering, wireless devices, biomedical implants, MEMS devices.
\end{IEEEbiography}

\begin{IEEEbiography}[{\includegraphics[width=1in,height=1.25in,clip,keepaspectratio]{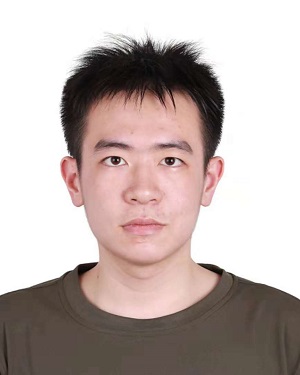}}]{Yan He} received the B.S degree in Electronic Science and Technology from the Zhejiang University, Hangzhou, China, in 2018. He is currently pursuing the Ph.D. degree in Electrical and Computer Engineering with the Rice University, Houston, TX, USA.

His current research interests include analog and mixed-signal integrated circuits design for power management and hardware security.
\end{IEEEbiography}

\begin{IEEEbiography}[{\includegraphics[width=1in,height=1.25in,clip,keepaspectratio]{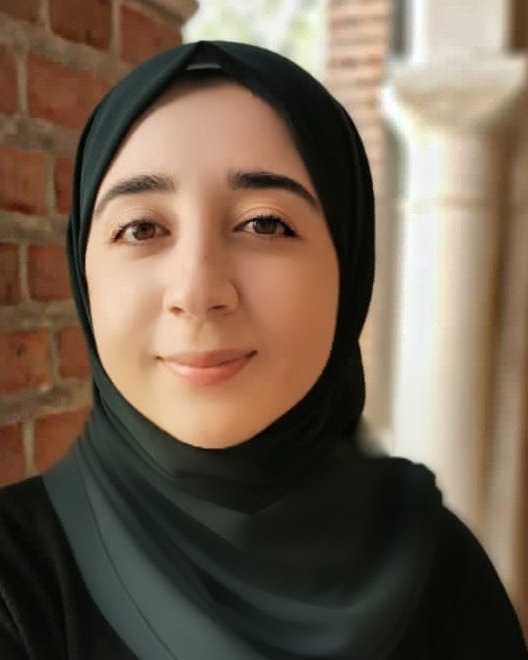}}]{Fatima Alrashdan} received a B.S and M.S degrees in Electrical Engineering from Jordan University of Science and Technology, Irbid, Jordan, in 2016 and 2018, respectively. She is currently working toward the Ph.D. degree in Electrical and Computer Engineering at Rice University, Houston, TX, USA. 

Her research interests include bioelectronics, wireless-implantable biomedical systems, power electronics and control systems for biomedical applications.   
\end{IEEEbiography}

\begin{IEEEbiography}[{\includegraphics[width=1in,height=1.25in,clip,keepaspectratio]{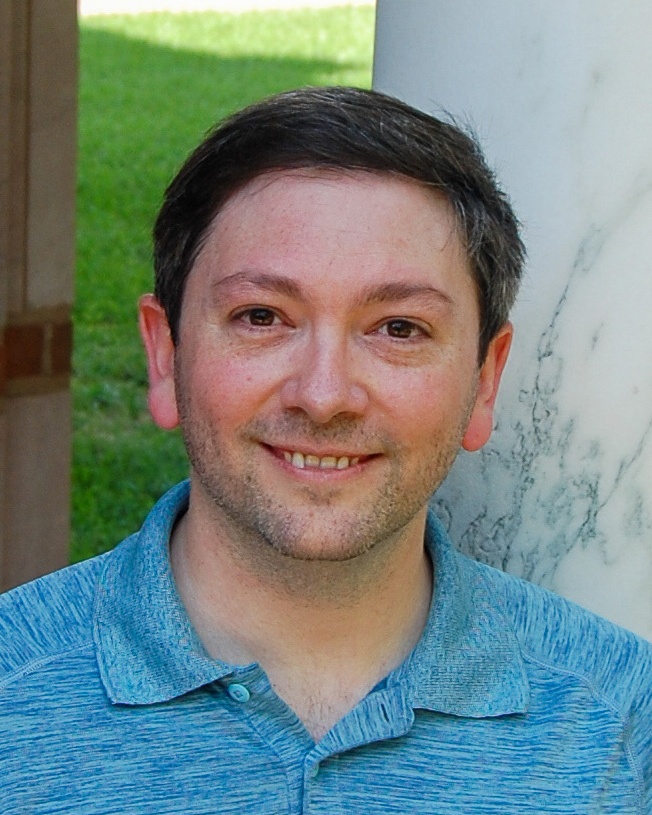}}]{Benjamin Avants} received a B.S. in Electrical Engineering, a B.S. in Computer Engineering, and second majors in Math and Physics from the University of Memphis, Tennessee in 2012. He then joined the Robinson Lab at Rice University, where he has worked as a research engineer ever since.  

Apart from electronics and programming, his interests also include 3D modeling, optics, imaging, and advanced fabrication on the nano, micro, and meso scale.  He enjoys developing researchers' dreams into realities whether it's in the clean room or a maker space.
\end{IEEEbiography}

\begin{IEEEbiography}[{\includegraphics[width=1in,height=1.25in,clip,keepaspectratio]{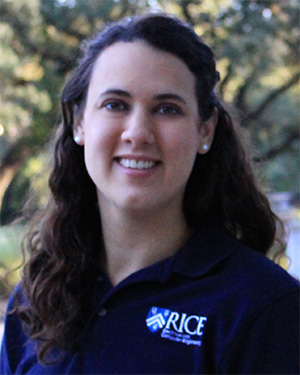}}]{Amanda Singer} received B.S. degree in Physics in 2014 from Saginaw Valley State University, University Center, MI, USA, and a M.A.Sc. degree in Applied Physics from Rice University, Houston, TX, USA, in 2018, and a Ph.D. degree in Applied Physics from Rice University, Houston TX, USA in 2021. She currently works as a postdoctoral fellow at Rice University.

Her current research interests include the development of wireless bioelectronics and the miniaturization of neural implants.
\end{IEEEbiography}

\begin{IEEEbiography}[{\includegraphics[width=1in,height=1.25in,clip,keepaspectratio]{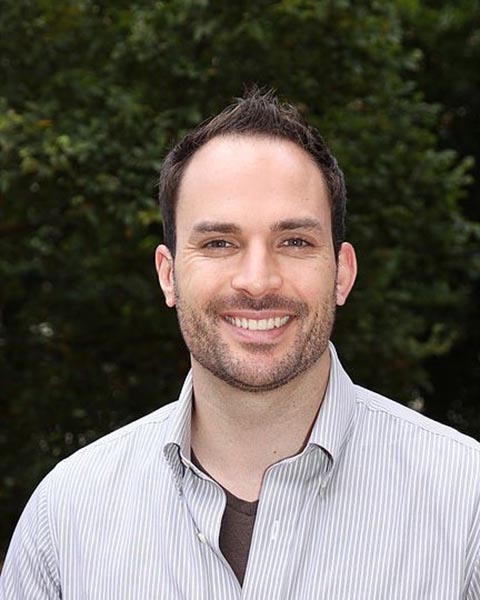}}]{Jacob Robinson}is an Associate Professor in Electrical and Computer Engineering and Bioengineering at Rice University and an Adjunct Assistant Professor in Neuroscience at Baylor College of Medicine. Dr. Robinson earned a B.S. in Physics from UCLA and a Ph.D. in Applied Physics from Cornell. Following his Ph. D., he worked as a postdoctoral fellow in the Chemistry Department at Harvard University. Dr. Robinson joined Rice University in 2012 where he currently works on nanoelectronic, nanophotonic, and nanomagnetic technologies to manipulate and measure brain activity. Dr. Robinson is an IEEE Senior member and currently a co-chair of the IEEE Brain Initiative. He is also the recipient of an NSF NeuroNex Innovation Award, DARPA Young Faculty Award, and Materials Today Rising Star Award.
\end{IEEEbiography}

\begin{IEEEbiography}[{\includegraphics[width=1in,height=1.25in,clip,keepaspectratio]{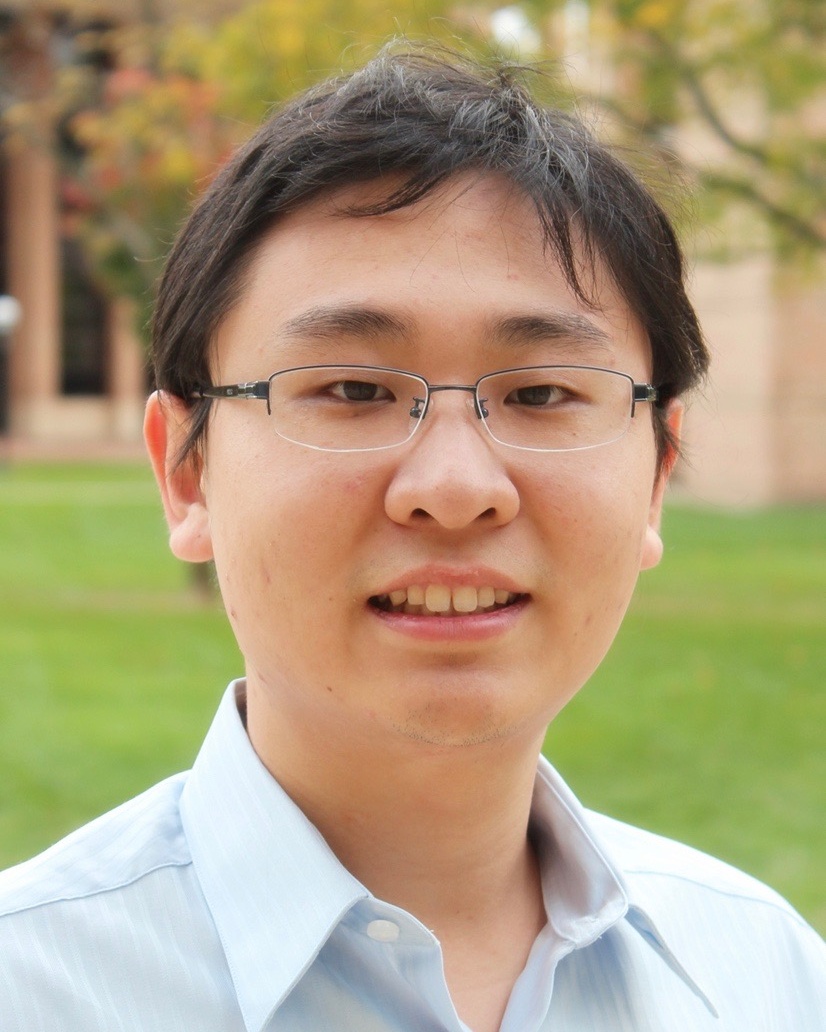}}]{Kaiyuan Yang} (S'13-M'17) received the B.S. degree in Electronic Engineering from Tsinghua University, Beijing, China, in 2012, and the Ph.D. degree in Electrical Engineering from the University of Michigan, Ann Arbor, MI, in 2017. His Ph.D. research was recognized with the 2016-2017 IEEE Solid-State Circuits Society (SSCS) Predoctoral Achievement Award. 

He is an Assistant Professor of Electrical and Computer Engineering at Rice University, Houston, TX. His research interests include digital and mixed-signal circuits for secure and low-power systems, hardware security, and circuit/system design with emerging devices. Dr. Yang received the Distinguished Paper Award at the 2016 IEEE International Symposium on Security and Privacy (Oakland), the Best Student Paper Award (1st place) at the 2015 IEEE International Symposium on Circuits and Systems (ISCAS), the Best Student Paper Award Finalist at the 2019 IEEE Custom Integrated Circuits Conference (CICC), and the 2016 Pwnie Most Innovative Research Award Finalist.
\end{IEEEbiography}



\vfill


\end{document}